\documentclass[11pt]{article}
\usepackage{geometry}
\usepackage{authblk}
\usepackage{latexsym,amsmath,amsfonts,amssymb,amsbsy,amsopn,amstext,amsthm}
\usepackage{epsfig,epstopdf,graphicx}
\usepackage[mathscr]{eucal}
\usepackage{multicol,makeidx}
\usepackage[monochrome]{xcolor}
\usepackage{multirow}
\usepackage{booktabs}
\usepackage{bm,mathrsfs}
\usepackage{float}
\usepackage{stackengine}
\usepackage{tablefootnote}
\usepackage{threeparttable}
\usepackage{mathdots}
\usepackage{tabu}
\usepackage{enumerate}
\usepackage[labelformat=simple]{subcaption}

\def\eq{\displaystyle\stackrel\triangle=}
\def\xra{\xrightarrow}

\def\d{\mathrm{d}}

\newcommand*\mcapinn[2]{\vcenter{\hbox{$\mathsurround=0pt
  \ifx\displaystyle#1\textstyle\else#1\fi\bigcap$}}}

\newcommand*\mcupinn[2]{\vcenter{\hbox{$\mathsurround=0pt
  \ifx\displaystyle#1\textstyle\else#1\fi\bigcup$}}}

\DeclareFontFamily{OT1}{pzc}{}
\DeclareFontShape{OT1}{pzc}{m}{it}{<-> s * [1.200] pzcmi7t}{}
\DeclareMathAlphabet{\mathpzc}{OT1}{pzc}{m}{it}

\graphicspath{{./images/}}

\newtheorem{theorem}{Theorem}[section]

\newtheorem{assum}{Assumption}
\newtheorem{thm}{Theorem}
\newtheorem{prop}{Proposition}
\newtheorem{lem}{Lemma}

\newtheorem{defn}{Definition}
\newtheorem{rmk}{Remark}

\renewenvironment{proof}[1][Proof]{\noindent\textbf{#1.} }{\hfill \rule{0.5em}{0.5em}}

\newenvironment{keyword}
{\par\small\noindent\textbf{\textit{Keywords---}}}
{\par}

\begin{document}

	\title{\textbf{Asymptotically Efficient Finite Impulse Response System Identification without Noise Knowledge}}
	\author{Zichen Liu and Biqiang Mu}

	\affil{State Key Laboratory of Mathematical Sciences, Academy of Mathematics and Systems Science, 
		 Chinese Academy of Sciences, Beijing 100190, China\\
	\texttt{(liuzichen@amss.ac.cn; bqmu@amss.ac.cn)}}

\date{}
\maketitle

	\begin{abstract}                          
Efficient identification of finite impulse response systems under unknown or heavy-tailed noise remains a  challenge in system identification. 
Classical estimators, including least squares (LS) and robust alternatives like least absolute deviations (LAD), are often sensitive to outliers or statistically inefficient when noise characteristics deviate from their underlying assumptions. To address this, we propose a novel class of  adaptive estimators that achieves asymptotic efficiency without requiring prior knowledge of the noise distribution. Our approach leverages kernel density estimation  to construct two nonparametric maximum likelihood-type estimators: the estimated maximum likelihood estimator (eMLE) and the estimated profiled MLE (epMLE). These estimators approximate the  likelihood using residuals from consistent initial estimates, avoiding parametric assumptions on the noise. We establish  theoretical guarantees showing that, under mild regularity conditions, both eMLE and epMLE are consistent, asymptotically normal, and  attain the Cramér-–Rao lower bound, thereby achieving  asymptotic efficiency. 
Additionally, we develop computationally efficient one-step Newton--Raphson algorithms that implement these estimators without sacrificing asymptotic efficiency. 
Extensive simulations across various noise distributions confirm that the proposed methods outperform classical LS and LAD estimators in both robustness and accuracy, often matching oracle MLE performance.

	\end{abstract}

    \begin{keyword}                           
	Linear system identification; Nonparametric maximum likelihood estimators;  Asymptotic efficiency; Kernel density estimation.
	\end{keyword}                             


	\section{Introduction}
	Efficient identification of linear dynamical systems is fundamental to modeling, prediction, and control, yet it remains significantly challenged by the presence of unknown system noise.
	Among linear model structures, the finite impulse response (FIR) system is widely employed due to its simplicity, finite parametrization, and interpretability \cite{Ljung1999}. An FIR system expresses the output as a linear convolution of past inputs with an unknown impulse response, reducing the identification task to a linear regression problem with temporally correlated regressors.
	
	Existing methods for FIR system identification inevitably rely on assumptions about the noise distribution, which may not hold in practice. 
	The most common is the Gaussian assumption, under which the classical least squares (LS) estimator \cite{marple1981,ljung1985,rao2009} is both consistent and asymptotically efficient. While LS remains consistent and asymptotically normal for noise with zero mean and finite variance \cite{sternby1977,nassiri1994}, it loses asymptotic efficiency under non-Gaussian distributions and can become inconsistent if the noise variance is infinite.

	A more principled route to efficiency is the maximum likelihood estimator (MLE), which assumes a known noise density and minimizes the negative log-likelihood. Under standard regularity conditions, the MLE achieves the Cramér–Rao lower bound (CRLB) \cite{self1987}. 
	However, this optimality depends critically on the correctness of the assumed density.
	A more flexible approach is to assume the noise belongs to a parametric family, allowing joint estimation of system and noise parameters within the M-estimation framework \cite{van2000}. 
	While this relaxes the need for an exact density, it still imposes structural constraints. The resulting estimator converges to a pseudo-parameter defined by minimizing the Kullback-Leibler divergence between the true density and the assumed parametric family \cite{white1982}. Efficiency is guaranteed only if the true density is within this family; otherwise, the estimator may be inefficient or even inconsistent.
	
	As a safeguard against noise misspecification, robust estimation methods, such as the Huber estimator \cite{Poljak1980,huber1981}  and least trimmed squares (LTS) estimators \cite{rousseeuw1985,bai2003,Mu2026}, as well as the least median of squares (LMS) 
	\cite{siegel1982},   
	typically guarantee consistency and asymptotic normality under a broader class of noise distributions. 
	The Huber estimator
	is consistent and asymptotically normal without
	requiring any moment conditions on the noise distribution. However, its performance is sensitive to a tuning parameter, and an inappropriate selection can lead to an inflated asymptotic variance \cite{bellec2025}.
	The LTS achieves $\sqrt{n}$-consistency and asymptotic normality, whereas the LMS converges at a slower rate of $\sqrt[3]n$, where $n$ is the sample size. 
	Despite their robustness, none of these methods attains asymptotic efficiency.

	In summary, existing FIR system identification methods cannot simultaneously handle diverse and unknown noise distributions while maintaining asymptotic efficiency. 
	Therefore, developing a consistent and efficient estimator that adapts to unknown noise without parametric or moment assumptions remains a significant open challenge.
	This motivates a distribution-free approach grounded in likelihood principles: ``Estimate the unknown noise density nonparametrically using kernel density estimation (KDE) \cite{silverman1986}, then plug it into the likelihood function."
	
	This idea is not new and has been investigated in statistics for various models under independent and identically distributed (i.i.d.) regressors, including linear models \cite{yao2013, chen2020, de2021}, nonlinear models \cite{yuan2007}, partial linear multivariate response models \cite{zhang2021}, nonparametric regression models \cite{linton2007, wang2024}, and additive models \cite{li2025}. A closely related approach is the minimum-entropy estimation \cite{wolsztynski2005}. 
	However, even with i.i.d. regressors, the asymptotic normality of these estimators for parametric models has not been rigorously established. 
	A different route, known as adaptive estimation, estimates the score function of the noise by KDE and then performs a one-step Newton--Raphson update from a preliminary estimator \cite{bickel1982, kreiss1987, koul1997}. Its asymptotic efficiency typically requires complex technical conditions, including symmetric noise, finite variance or higher-order moments, and sample splitting. 
	In contrast, this idea has received little attention in system identification. To the best of our knowledge, the only related work is \cite{bercu2008}, where KDE is used to estimate the noise density and to test the goodness-of-fit in adaptive tracking, while the parameters are still estimated by the extended least squares method. 
	Direct extension to FIR system identification is non-trivial due to: (i) temporal dependence in regressors, (ii) lack of principled bandwidth selection, and (iii) non-convex optimization with the risk of converging to a poor local minima.
	
	To address this gap, this paper introduces two nonparametric MLEs: the estimated MLE (eMLE) and the estimated profiled MLE (epMLE). These methods approximate the likelihood via KDE using residuals from an initial consistent estimator, without imposing parametric or moment assumptions on the noise. 
	To the best of our knowledge, we give the first rigorous proof of their asymptotic normality.
	Moreover, to improve computational efficiency, we propose one-step Newton--Raphson algorithms that maintain asymptotic efficiency while significantly reducing computational costs.

	This paper makes the following key contributions:
	\begin{enumerate}[(i)]
		\item We propose a novel class of nonparametric MLEs for FIR system identification, which does not require prior knowledge of the noise distribution;
		
		\item We prove consistency and asymptotic efficiency of the proposed estimators and their resulting one-step Newton--Raphson algorithms for temporally correlated regressors of FIR systems. This fills the theoretical gap left even in the i.i.d. setting;
		
		\item We provide a detailed discussion on obtaining a $\sqrt{n}$-consistent initial estimator, ensuring reliable convergence and avoiding poor local optima in the non-convex optimization;
		
		\item We introduce a robust $K$-fold cross-validation procedure for jointly selecting the bandwidth and initial parameters, overcoming limitations of empirical rules or loss functions sensitive to heavy-tailed noise.

	\end{enumerate}

	The remainder of this paper is organized as follows. Section \ref{sec:2} introduces the FIR system identification problem and reviews key preliminaries, including LS, MLE, and  KDE. Section \ref{sec:3} introduces the proposed eMLE and epMLE methods and establishes their asymptotic efficiency. Section \ref{sec:5} presents the corresponding computationally efficient one-step Newton--Raphson algorithms, proves their asymptotic efficiency, and   discusses practical considerations such as bandwidth selection and initialization.  Section \ref{sec:6} provides extensive simulation studies under various noise conditions, demonstrating the robustness and superior accuracy of the proposed methods  compared to existing alternatives.
	Finally, Section \ref{sec:7} concludes the paper.
	
	\textbf{Notation.}
	We use the following notation throughout the paper. 
	The symbols  \(\mathbb{P}\) and \(\mathbb{E}\) denote the probability of an event and the expectation operator with respect to the underlying probability measure, respectively.
	For two deterministic sequences $\{a_n\}$ and $\{b_n\}$, $a_n=O(b_n)$ means that there exist a constant $C>0$ and a positive integer $N$ such that $|a_n/b_n|\leq C$ for all $n\geq N$.
	For a sequence of random variables \(\{X_n\}\), (i)  \(X_n = O_p(1)\)  means that \(X_n\) is bounded in probability (i.e.,  for any $\epsilon> 0$, there exist a constant $C>0$ and a positive integer $n_0$ such that $\mathbb{P}(|X_n|>C)<\epsilon$ for all $n>n_0$); (ii) \(X_n = o_p(1)\) (or \(X_n  \xra{p} 0\)) indicates that \(X_n\) converges to zero in probability (i.e., 
	for any $\epsilon> 0$,  there holds that $\lim_{n\xra{}\infty}\mathbb{P}(|X_n|>\epsilon)=0$); 
	(iii) $X_n\xra{d}\mathcal{N}(0,\sigma^2)$ denotes that $X_n$ converges in distribution to a Gaussian random variable with mean zero and variance $\sigma^2>0$; (iv) $\mathbb{E}[X_1|X_2]$ stands for conditional expectation of $X_1$ given random variable $X_2$.
	For a vector $a$, $\|a\|$ means the $\ell_2$-norm, and $a^{\otimes2}$ represents its outer-product $aa^T$.
	For an $r$-times differentiable function $f(\cdot)$,  $f^{(i)}(\cdot)$ denotes its $i$-th order  derivative for $1\leq i\leq r$.

	\section{Preliminaries}\label{sec:2}

	\subsection{Finite impulse response systems}
	Consider the single-input single-output FIR system described by 
	\begin{subequations}
		\label{fir}
		\begin{align}
			&y_k = G(q,\beta_0)u_k +\varepsilon_k,~~k=1,\cdots,n,\\
			&G(q,\beta_0)=\sum_{i=1}^pg_i^0q^{-i},~\beta_0=[g_1^0,g_2^0,\cdots,g_p^0]^T,
		\end{align}
	\end{subequations}
	where $p$ is the system order, $\beta_0$ is the true parameter, and
	$y_k,u_k$, and $\varepsilon_{k}$ are the output, input and noise, respectively.
	The FIR system has a linear regression form:
	\begin{align}
		y_k=x_k^T\beta_0+\varepsilon_k, ~
		x_k=[u_{k-1},u_{k-2},\cdots,u_{k-p}]^T.
		\label{split}
	\end{align}
	We rewrite the FIR system \eqref{split} in the vector form:
	\begin{align}
		y=X\beta_0+\varepsilon ,
	\end{align}
	where $y=[y_{1},y_{2},\cdots,y_n]^T,$ $X=[x_{1},x_{2},\cdots,x_{n}]^T,$ and $\varepsilon=[\varepsilon_{1},\varepsilon_{2},\cdots,\varepsilon_{n}]^T.$
	
	The goal of the FIR system identification is to estimate the true parameter $\beta_0$ of the FIR system \eqref{fir} as accurately as possible using the data $\{u_k,y_k,k=1,\ldots,n\}$.
	
	We first give the assumption on the input throughout the paper.
	\begin{assum}\label{asp1}
		The input $\{u_k$\} is a sequence of i.i.d. random variables with mean zero, variance $\sigma^2_u>0$, and finite sixth-order moment.
	\end{assum}
	
	\subsection{Least squares estimators}
	One prevalent method for identifying the true parameter $\beta_0$ is  the  LS estimator:
	\begin{align}
		\widehat{\beta}^{LS}\eq\mathop{\arg\min}_{\beta\in\mathbb{R}^p}\| y-X\beta\|^{2}
		=(X^{T}X)^{-1}X^{T}y.
		\label{ls}
	\end{align}
	The performance of the LS estimator depends critically on the noise distribution. Under the  i.i.d. Gaussian noise, it is both unbiased and asymptotically efficient. However, under infinite-variance noise, the estimator becomes inconsistent and may produce arbitrarily poor results.
	
	\subsection{Classical maximum likelihood estimators}
	Conversely, the MLE can overcome a key limitation of the LS estimator that it is  restricted to finite-variance or Gaussian noise. In particular, the MLE remains well-defined, and can be asymptotically efficient for a broad class of noise distributions, including those with heavy tails or infinite variance, provided the true density  is known. 
	Assume the noise is independently generated from a common distribution with probability density function $f_o(\cdot)$.
	Then, the MLE of $\beta_0$ is given by
	\begin{equation}\label{mle}
		\begin{aligned}
			\widehat{\beta}^{M}\eq\mathop{\arg\max}_{\beta\in\mathbb{R}^{p}}\prod_{k=1}^{n} f_o(y_k-x_k^T\beta)  =\mathop{\arg\max}_{\beta\in\mathbb{R}^{p}}\frac1n\sum_{k=1}^{n}\log f_o(y_k-x_k^T\beta).
		\end{aligned}
	\end{equation}
	Under regularity conditions, the MLE is consistent and asymptotically normal \cite{Ljung1999,van2000},
	\begin{subequations}
		\begin{align}
			\sqrt{n}(\widehat{\beta}^{M}-\beta_0)\xrightarrow{d}\mathcal{N}({0},\Sigma^{-1}),
		\end{align}
		where
		\begin{align}
			\Sigma \eq \mathbb{E}\left[\left(f_o^{(1)}(\varepsilon_k)\Big/f_o(\varepsilon_k)\right)^2\right]\sigma_u^2 I_p,
		\end{align}
	\end{subequations}
	and $I_p$ is the $p\times p$ identity matrix.
	This indicates that the MLE \eqref{mle} achieves the CRLB and has the smallest asymptotic variance among all unbiased estimators \cite{Ljung1999,van2000}, but crucially depends on the true density  
	$f_o(\cdot)$.

	\subsection{Nonparametric kernel density estimation}
	In many practical settings, the true noise density  is unknown, rendering the   MLE in \eqref{mle} infeasible. To address this, we propose replacing  $f_o(\cdot)$ with a nonparametric kernel density estimator (KDE) \cite{Wand1994,Greblicki2008}. This substitution serves as a key technical device to eliminate the need for parametric assumptions on the noise distribution, while still preserving the efficiency and interpretability of likelihood-based estimation.
	
	For the following nonparametric MLEs, we first introduce the nonparametric KDE for estimating $f_o(\cdot)$ directly from noise samples. 
	Given i.i.d. noise samples $\{ \varepsilon_1, \varepsilon_2, \dots, \varepsilon_n \}$ drawn from $f_o$,  the KDE of $f_o(\cdot)$ at point $t$ is defined as
	\begin{align}\label{kde}
		\widetilde{f}_h(t) \eq \frac{1}{n h} \sum_{k=1}^{n} K\left( \frac{t-\varepsilon_k}{h} \right),
	\end{align}
	where $K(\cdot)$ is a kernel function, typically  a smooth, symmetric, unimodal probability density function (e.g., Gaussian kernel, Epanechnikov kernel, cosine kernel, uniform kernel, triangular kernel, etc.), and $h$ is the bandwidth parameter controlling the bias-variance trade-off.
	
	The mean squared error of the KDE decomposes into  bias and variance components. For any given $t$, the bias and variance are
	$\mathbb{E}[\widetilde{f}_h(t)]-f_o(t)$ and $\mathbb{E}\big[\widetilde{f}_h(t)
	-\mathbb{E}\widetilde{f}_h(t)
	\big]^2$, respectively. 
	Under standard smoothness assumptions, the orders of the bias and variance are
	\begin{align*}
		\left|\mathbb{E}[\widetilde{f}_h(t)]-f_o(t)\right|=O(h^2),~~\mathbb{E}\big[\widetilde{f}_h(t)
		-\mathbb{E}\widetilde{f}_h(t)
		\big]^2=O\Big(\frac{1}{nh}\Big).
	\end{align*}
	Moreover, when the noise density function has higher-order smoothness, we can utilize the higher-order kernel technique to reduce the bias of the KDE \cite{jones1997}. We formalize this as the following definition.
	\begin{defn} (Higher-order kernel)
		A kernel function $K(\cdot)$ is said to be a kernel of order $q\  (q\geq 1)$ if
		\begin{align*}
			\int \!K(x)x^{j}{\rm d}x\!=\!0 ~\text{for} ~ j=1,\ldots,2q\!-\!1,  ~
			\int K(x)x^{2q}{\rm d}x<\infty.
		\end{align*}
	\end{defn}
	If the density function has a bounded $2q$-th order derivative, then using a $q$-th order kernel leads to an improved bias such that
	$$
	\left|\mathbb{E}[\widetilde{f}_h(x)]-f_o(x)\right|=O(h^{2q}),
	$$
	while the order of variance remains unchanged.
	
	The bandwidth $h$ plays a pivotal role in balancing bias and variance of the KDE:
	(i) A small $h$ leads to low bias but large variance (undersmoothing/overfitting);
	(ii) A large $h$ yields low variance but large bias (oversmoothing). 
	Although Silverman’s rule of thumb offers a quick, Gaussian-based bandwidth choice, it lacks adaptivity to non-Gaussian or multimodal densities. Consequently, data-driven bandwidth selection methods such as cross-validation methods are generally preferred for adaptive bandwidth selection in practice.
	
	\section{Nonparametric maximum likelihood estimators}\label{sec:3}
	
	In this section, we introduce two nonparametric MLEs, the estimated MLE (eMLE) and the estimated profiled MLE (epMLE), which replace the unknown noise density \(f_o\) in the likelihood with the KDE constructed from (pseudo) noise samples. We establish that both nonparametric MLEs are consistent and asymptotically efficient, achieving the CRLB without requiring prior knowledge of the noise distribution.

	\subsection{Two nonparametric  MLEs}
	
	In this subsection, we present two nonparametric MLEs: the eMLE relying on an initial consistent estimator of the true parameter and the  epMLE without using an initial estimator.
	
	\subsubsection{Estimated maximum likelihood estimators}
	
	Although the KDE can be used to construct a nonparametric estimate for the likelihood function in \eqref{mle}, we do not observe direct samples from the noise distribution. To address this, we leverage an initial estimator of the true  parameters to generate pseudo-noise samples via residuals.
	
	Specifically, we estimate the noise samples
	$\varepsilon_k$ by the residual 
	\begin{equation}\label{epshatbeta}
		\widehat{\varepsilon}_k\eq y_k-x_k^T\widehat{\beta}^{(0)},~k=1,\cdots,n,
	\end{equation}
	where $\widehat{\beta}^{(0)}$ is an initial estimator of $\beta_{0}$. 
	The choice of $\widehat{\beta}^{(0)}$ will be discussed in Section \ref{sec5.3}. 
	Using these estimated residuals, we construct the estimator of noise density $f_o(\cdot)$:
	\begin{align}\label{kdefunc}
		\widehat{f}_{h}(t)\eq\frac{1}{nh}\sum_{k=1}^{n}K\left(\frac{t-\widehat{\varepsilon}_k}{h}\right).
	\end{align}
	
	The eMLE of $\beta_0$  is then   defined as the maximizer of the estimated likelihood function: 
	\begin{equation}\label{eM}
		\begin{aligned}
			\widehat{\beta}^{eM}\!\eq\mathop{\arg\max}_{\beta\in\mathbb{R}^{p}}\prod_{k=1}^{n} \!\widehat{f}_{h}\!\left(y_k-x_k^T\beta\right)\!=\mathop{\arg\max}_{\beta\in\mathbb{R}^{p}}\widehat{L}_{n}^{eM}(\beta),
		\end{aligned}
	\end{equation}
	where 
	$$
	\widehat{L}_n^{eM}(\beta)\eq\frac{1}{n}\sum_{k=1}^{n}\log \widehat{f}_{h}(y_k-x_k^T\beta).
	$$
	An alternative is to  use a leave-one-out version of the estimated density function to mitigate potential bias from using the same residuals in both density estimation and likelihood evaluation. This yields the leave-one-out eMLE: 
	\begin{align*}
		\widehat{\beta}_{lo}^{eM}\! \eq\!\mathop{\arg\max}_{\beta\in\mathbb{R}^{p}}\!\sum_{k=1}^{n}\log\!\Bigg\{\!\frac{1}{(n\!-\!1)h}\!\sum_{j\neq k}^{n}\!K\!\Big(\frac{(y_k-x_k^T\beta)-\widehat{\varepsilon}_j}{h}\Big)\!\!\Bigg\}.
	\end{align*}
	Under standard regularity conditions, the difference between the standard eMLE and its leave-one-out counterpart is asymptotically negligible. Consequently, we focus our theoretical analysis and numerical experiments on the eMLE throughout the paper.
	
	\subsubsection{Estimated profiled maximum likelihood estimators}
	
	In contrast to the eMLE, the epMLE directly utilizes the residuals
	\begin{equation}\label{epsbeta}
		\varepsilon_k(\beta)\eq y_k-x_k^T\beta,~k=1,\cdots,n,
	\end{equation}
	which depend on the parameter vector $\beta$ that is simultaneously optimized in the estimation procedure.
	Accordingly, we define a kernel density estimator that explicitly depends on $\beta$: 
	\begin{align}
		\widehat{f}_{h}(t,\beta)\eq\frac{1}{nh}\sum_{k=1}^{n}K\Big(\frac{t-\varepsilon_k(\beta)}{h}\Big).\label{fbetat}
	\end{align}
	The epMLE is then defined by maximizing the estimated likelihood function with respect to $\beta$:
	\begin{equation}
		\begin{aligned}\label{epM}
			\widehat{\beta}^{epM}&\!\eq\mathop{\arg\max}_{\beta\in\mathbb{R}^{p}}\prod_{k=1}^{n} \widehat{f}_{h}\!\left(y_k-x_k^T\beta,\beta\right)=\mathop{\arg\max}_{\beta\in\mathbb{R}^{p}}\widehat{L}_n^{epM}(\beta),
		\end{aligned}
	\end{equation}
	where
	$$
	\widehat{L}_{n}^{epM}(\beta)\eq\frac{1}{n}\sum_{k=1}^{n}\log \widehat{f}_h(y_k-x_k^T\beta,\beta).
	$$
	As with the eMLE, one may also consider a leave-one-out variant to reduce potential bias:
	\begin{align*}
		\widehat{\beta}_{lo}^{epM}\!\eq\mathop{\arg\max}_{\beta\in\mathbb{R}^{p}}\sum_{k=1}^{n}\!\log\!\Bigg\{\!\frac{1}{(n-1)h}\sum_{j\neq k}K\!\Big(\frac{\varepsilon_k(\beta)-\varepsilon_j(\beta)}{h}\Big)\!\!\Bigg\}.
	\end{align*}
	Under standard regularity conditions, the asymptotic properties of the epMLE and its leave-one-out version are equivalent. Thus, for simplicity, we focus our analysis on the epMLE.

	\subsection{Statistical efficiency}

	In this subsection, we show that both the eMLE and epMLE achieve the same asymptotic covariance matrix as that of the MLE. As a result, these estimators are asymptotically efficient  without requiring prior knowledge of the true noise density $f_o$.

	\subsubsection{Asymptotic efficiency of eMLE}
	Before presenting the asymptotic properties of the eMLE, we introduce the assumptions concerning the noise,  kernel function and its bandwidth,  initial estimator.

	\begin{assum}\label{asp2}
		
		\begin{enumerate}[(i)]
			\item The noise $\{\varepsilon_k\}$ is a sequence of i.i.d. random variables with the density function $f_o(\cdot)$ and is independent of the input $\{u_k\}$. 
			\item $f_o(\cdot)$ has bounded derivatives up to order four.
			\item $\mathbb{E}\big[\big(f_o^{(1)}(\varepsilon_k)\big/f_o(\varepsilon_k)\big)^2\big]>0$ and $\inf_{t\in\mathcal{T}}f_o(t)>0$, where $\mathcal{T}$ is  the support of $f_o(t)$.
		\end{enumerate}
	\end{assum}
	
	\begin{assum}\label{asp3}
		\begin{enumerate}[(i)]
			\item   The limit function of $\widehat{L}_{n}^{eM}(\beta)$,
			\begin{align}
				L_{eM}(\beta)=\mathbb{E}[\log f_o(\varepsilon_k(\beta))]\label{mf},
			\end{align} has continuous positive definite Hessian matrix, and is well-separated around $\beta_0$, that is, for any $\delta>0$,
			$$
			\sup_{\beta:\Vert\beta-\beta_0\Vert\geq\delta}L_{eM}(\beta)<L_{eM}(\beta_0).
			$$
			\item $\mathbb{E}[\sup\limits_{\beta\in{\Theta}}|\log f_o(\varepsilon_k(\beta))|^{\gamma}]<\infty$ for some $\gamma>1$, where $\Theta$ is a compact set including $\beta_0$.
		\end{enumerate}
	\end{assum}
	In Assumption \ref{asp2}-(\romannumeral3), we assume the density has a positive lower bound on its support, i.e.,  $\inf_{t\in\mathcal{T}}f_o(t)>0$, which is a technical condition introduced for the convenience of theoretical proof. This condition can be removed by employing a trimmed version of the objective function $\widehat{L}_n^{eM}(\beta)$ \cite{yao2013}.
	Assumption~\ref{asp3}-(\romannumeral1) ensures the identifiability of the true parameter vector \(\beta_0\). This is a standard condition in M-estimation theory; see, for example, \cite[Chapter 5]{van2000}.
	Assumption~\ref{asp3}-(\romannumeral2) imposes an adaptive moment condition on the noise distribution, which is crucial for establishing consistency of the MLE (see \cite[Section 5.2.1]{van2000}). Under Gaussian noise, this condition requires a finite \(2\gamma\)-th moment; for Laplace noise, it requires a finite \(\gamma\)-th moment. Notably, the condition also holds for the Cauchy distribution, since the expectation of logarithmic-polynomial functions is finite under Cauchy noise despite its heavy tails and undefined absolute mean.

	\begin{assum}\label{asp4}
		The initial estimator $\widehat{\beta}^{(0)}$ is   $\sqrt{n}$-consistent, i.e., $\beta_0-\widehat{\beta}^{(0)}=O_p({1/\sqrt{n}}).$
	\end{assum}

	\begin{assum}\label{asp5}
		
		\begin{enumerate}[(i)]
			\item The kernel function $K(\cdot)$ is symmetric about 0, and
			has continuous derivatives up to the fourth order;
			\item For $r=1,2,3,4$, $\sup_{x}|K^{(r)}(x)|<\infty$ and the integrals  associated with the derivatives $K^{(r)}(\cdot)$ are finite, i.e., 
			$$
			\int \big|K^{(r)}(x)\big|^{s}|x|^{t}{\rm d}x <\infty, ~s=1,2,~t=0,1,2.
			$$ Moreover, both $K^{(1)}(x)$ and $K^{(2)}(x)$ are bounded variation functions;
			
			\item The bandwidth satisfies $h=O(n^{-\alpha})$ for some $\alpha$ satisfying $1/8<\alpha<1/6$.
		\end{enumerate}
		
	\end{assum}

	\begin{thm}\label{thm1}
		Suppose Assumptions \ref{asp1}--\ref{asp5} hold.
		Then the eMLE estimator $\widehat{\beta}^{eM}$ is consistent and   asymptotically normal, satisfying
		$
		\sqrt{n}(\widehat{\beta}^{eM}-\beta_0)\xrightarrow[]{d}\mathcal{N}(0,\Sigma^{-1}).
		$
		Moreover, the asymptotic covariance matrix $\Sigma^{-1}$ achieves the CRLB, showing that $\widehat{\beta}^{eM}$ is asymptotically efficient.
	\end{thm}
	
	\subsubsection{Asymptotic efficiency of epMLE}
	To establish the asymptotic efficiency of the epMLE \(\widehat{\beta}^{epM}\), we require an additional condition on the kernel function and the noise density, as well as a counterpart to Assumption~\ref{asp3}-(\romannumeral1) that ensures identifiability in the profiled likelihood setting.
	
	To reduce the bias in the kernel density estimation, we adopt the higher-order kernel in the asymptotic results of epMLE, which is given as follows.
	\begin{assum}\label{asp6}
		The kernel $K(\cdot)$ is a higher-order kernel of order $q\geq3$, and $f_o$ is $2q+1$-times continuously differentiable.
	\end{assum}
	
	\begin{assum}\label{asp7}
		\begin{enumerate}[(i)]
			\item 
			The limit function of $\widehat{L}_{n}^{epM}(\beta)$,
			\begin{align*}
				L_{epM}(\beta)\!\eq\!\mathbb{E}\big[\!\log\mathbb{E}[f_o\!\left(\widetilde{y}_k\!-\!\widetilde{x}_k^T\!\beta\!-\!x_k^T\!(\beta_0\!-\!\beta)\right)\!|\widetilde{x}_k,\widetilde{y}_k]\big]
			\end{align*}
			has continuous positive definite Hessian matrix, and is well-separated around $\beta_0$, where $\widetilde{x}_k$  and $\widetilde{y}_k$ are independent copies of $x_k$ and $y_k$, respectively. 
			
			\item 
			$
			\mathbb{E}\Big[\sup\limits_{\beta\in\Theta}\Big|\log\mathbb{E}\!\left[f_o\!\left(\widetilde{y}-\widetilde{x}^T\!\beta-x^T\!(\beta_0\!-\!\beta)\right)\!\big|\widetilde{x},\widetilde{y}\right]\!\Big|^{\gamma}\Big]\!<\!\infty,
			$ for some $\gamma>1$,
			where $\Theta$ is a compact set including $\beta_0$.
		\end{enumerate}

	\end{assum}

	\begin{thm}\label{thm2}
		Suppose Assumptions \ref{asp1}--\ref{asp2} and \ref{asp5}--\ref{asp7} hold. Then the epMLE estimator $\widehat{\beta}^{epM}$ is consistent and asymptotically efficient, satisfying
		$
		\sqrt{n}(\widehat{\beta}^{epM}-\beta_0)\xrightarrow[]{d}\mathcal{N}(0,\Sigma^{-1}).
		$
	\end{thm}
	Theorems~\ref{thm1} and~\ref{thm2} demonstrate that, by leveraging KDE, both the eMLE and epMLE are asymptotically efficient: they achieve the same asymptotic variance as the MLE without utilizing the density of the true noise distribution.

	\section{Practical issues}\label{sec:5}
	In this section, we discuss practical considerations for implementing the eMLE and the epMLE, focusing on three key aspects: computationally efficient implementation,  initialization, and joint selection of bandwidth and  initial estimators.

	\subsection{Efficient Computation}\label{sec:4a}
	
	The computation of both the eMLE and epMLE involves solving a non-convex optimization problem, for which standard optimization algorithms often struggle to locate the global optimum without a reliable initial guess.  
	
	To ensure both theoretical guarantees and computational tractability, we propose a one-step Newton-Raphson algorithm initialized by a $\sqrt{n}$-consistent preliminary estimator $\widehat{\beta}^{(0)}$ \cite{Lehmann1998}. 
	Here, the consistent initial estimator can be chosen between LS and LAD estimators as discussed in Section \ref{sec:5.1}.
	This approach yields estimators that are asymptotically equivalent to the full eMLE and epMLE, while avoiding the difficulties associated with direct maximization of the non-convex objective functions.
	
	The one-step eMLE (oeMLE) estimator is given by:
	\begin{subequations}
		\begin{align}
			\label{oemle}
			\widehat{\beta}_{o}^{eM}=\widehat{\beta}^{(0)}-\widehat{H}_{eM}(\widehat{\beta}^{(0)})^{-1}\widehat{S}_{eM}(\widehat{\beta}^{(0)}),
		\end{align}
		where
		\begingroup
		\allowdisplaybreaks
		\begin{align}
			\label{sem}
			&\widehat{S}_{eM}(\beta)\eq\frac{\partial[- \widehat{L}_{n}^{eM}(\beta)]}{\partial\beta}=\frac{1}{n}\sum_{k=1}^{n}\frac{\widehat{f}_{h}^{(1)}(\varepsilon_k(\beta))}{\widehat{f}_{h}(\varepsilon_k(\beta))}x_k,\\
			\label{HEM}
			&\widehat{H}_{eM}(\beta)\eq
			\frac{\partial^2 [-\widehat{L}_{n}^{eM}(\beta)]}{\partial\beta\partial\beta^T}  =\frac{1}{n}\sum_{k=1}^{n}\Big(\frac{\widehat{f}_h^{(1)}(\varepsilon_k(\beta))}{\widehat{f}_h(\varepsilon_k(\beta))}\Big)^2x_kx_k^T-\frac{1}{n}\sum_{k=1}^{n}\frac{\widehat{f}_h^{(2)}(\varepsilon_k(\beta))}{\widehat{f}_h(\varepsilon_k(\beta))}x_kx_k^T,
		\end{align}
		\endgroup
	\end{subequations}
	and $\varepsilon_k(\beta)$ is defined in \eqref{epsbeta}, and for $r=1,2$, $\widehat{f}_{h}^{(r)}(t)$ is the $r$-th order derivative of $\widehat{f}_h(t)$, defined as
	\begin{align}
		\widehat{f}_{h}^{(r)}(t)=\frac{1}{nh^{r+1}}\sum_{k=1}^{n}K^{(r)}\Big(\frac{t-\widehat{\varepsilon}_{k}}{h}\Big).\label{fr}
	\end{align}
	
	The one-step epMLE (oepMLE) estimator is given by:
	\begin{subequations}
		\begin{align}
			\label{oepmle}
			\widehat{\beta}_{o}^{epM}=\widehat{\beta}^{(0)}-\widehat{H}_{epM}(\widehat{\beta}^{(0)})^{-1}\widehat{S}_{epM}(\widehat{\beta}^{(0)}),
		\end{align}
		where
		\begingroup
		\allowdisplaybreaks
		\begin{align}
			\label{sepm}
			&\widehat{S}_{epM}(\beta)\eq\frac{\partial[- \widehat{L}_{n}^{epM}(\beta)]}{\partial\beta} =\frac{1}{nh}\sum\limits_{k=1}^{n}\frac{\sum_{j=1}^{n}K^{(1)}\left(\frac{\varepsilon_{k}(\beta)-\varepsilon_{j}(\beta)}{h}\right)(x_k-x_j)}{\sum_{j=1}^{n}K\left(\frac{\varepsilon_{k}(\beta)-\varepsilon_{j}(\beta)}{h}\right)},\\
			\label{hepm}
			&\widehat{H}_{epM}(\beta)\eq\frac{\partial^2 [-\widehat{L}_{n}^{epM}(\beta)]}{\partial\beta\partial\beta^T} \\
			\nonumber
			&=\frac{1}{nh^2}\sum_{k=1}^{n}\left\{\frac{\left(\sum\limits_{j=1}^{n}K^{(1)}\left(\frac{\varepsilon_k(\beta)-\varepsilon_j(\beta)}{h}\right)(x_k\!-\!x_j)\right)^{\otimes2}}{\left(\sum\limits_{s=1}^{n}K\left(\frac{\varepsilon_{k}(\beta)-\varepsilon_s(\beta)}{h}\right)\right)^2} -\frac{\sum\limits_{j=1}^{n}K^{(2)}\left(\frac{\varepsilon_k(\beta)-\varepsilon_j(\beta)}{h}\right)(x_j\!-\!x_k)(x_j\!-\!x_k)^T}{\sum\limits_{s=1}^{n}K\left(\frac{\varepsilon_k(\beta)-\varepsilon_s(\beta)}{h}\right)}\right\}.
		\end{align}
		\endgroup
	\end{subequations}
	\begin{thm}\label{thm3}
		Suppose that Assumptions \ref{asp1}--\ref{asp5}
		hold. Then the one-step eMLE estimator enjoys the same asymptotic properties as $\widehat{\beta}^{eM}$, i.e.,
		$
		\sqrt{n}(\widehat{\beta}_{o}^{eM}-\beta_0)\xrightarrow[]{d}\mathcal{N}(0,\Sigma^{-1}).
		$
	\end{thm}
	
	\begin{thm}\label{thm4}
		Suppose Assumptions \ref{asp1}--\ref{asp2} and \ref{asp4}--\ref{asp7} hold. Then the one-step epMLE estimator enjoys the same asymptotic properties as $\widehat{\beta}^{epM}$, i.e.,
		$
		\sqrt{n}(\widehat{\beta}_{o}^{epM}-\beta_0)\xrightarrow[]{d}\mathcal{N}(0,\Sigma^{-1}).
		$
	\end{thm}
	
	\subsection{Discussion on initial estimators}\label{sec:5.1}
	
	The LS estimator \eqref{ls} is a common initializer for the eMLE. However, its consistency relies on the noise having finite variance. For heavy-tailed distributions like Student's 
	\(t(2)\), the LS estimator is inconsistent, violating Assumption \ref{asp5} and providing a poor starting value that can severely degrade the eMLE's performance.

	To address this, we recommend the least absolute deviation (LAD) estimator as a more robust alternative. It is defined as  
	\[
	\widehat{\beta}^{ LAD} = \underset{\beta}{\arg\min} \; \frac{1}{n} \sum_{i=1}^{n} |y_i - x_i^T \beta|.
	\]  
	Unlike LS, the LAD estimator remains \(\sqrt{n}\)-consistent under much weaker conditions: it only requires that the noise distribution has a zero median and a  continuous,  positive density \(f_o\) in a neighborhood of zero.
	This attractive property of the LAD estimator is formalized in the following result.
	
	\begin{prop}\cite{Bassett1978}
		\label{refthm:1}
		Suppose the noise \(\{\varepsilon_i\} \) is a sequence of i.i.d. random variables with  median zero and a continuous density that is positive in a neighborhood of zero. Then  
		\[
		\sqrt{n}\,(\widehat{\beta}^{  LAD} - \beta_0) \xrightarrow{d} \mathcal{N}\big(0, \Sigma_{  LAD}^{-1}\big),
		\]  
		where  
		$
		\Sigma_{ LAD} \eq 4 f_o^2(0) \, \sigma_u^2 I_p.
		$  
	\end{prop}
	A detailed treatment of the asymptotic theory of the LAD estimator can be found in \cite{Bassett1978}. 
	
	\begin{rmk}
		Generally, we do not require the noise to have zero median or zero mean; an intercept term can be incorporated into LS and LAD formulations and estimated jointly, which preserves the $\sqrt{n}$-consistency of initial estimators.
	\end{rmk}
	
	\begin{rmk}
		When the noise is suspected to be heavy-tailed, we recommend using the LAD estimator rather than the LS estimator as the initial estimator since the LAD estimator is more robust than LS and retains $\sqrt{n}$-consistency without imposing moment conditions on the noise.
	\end{rmk}
	
	\begin{rmk}
		In addition to the LAD estimator, the Huber estimator, defined through the loss function \begin{align}
			\label{huber}
			g(t) =   \begin{cases}
				\frac{1}{2} t^2, & \text{if } |t| \leq \delta, \\
				\delta \left(|t| - \frac{1}{2} \delta\right), & \text{if } |t| > \delta
			\end{cases}
		\end{align} 
		with tuning parameter \(0\leq \delta <\infty\), is also \(\sqrt{n}\)-consistent and does not require moment conditions on the noise distribution. Moreover, it offers a favorable trade-off between efficiency under light-tailed errors and robustness under heavy-tailed or contaminated data.
		However, in practice, to avoid the need to tune the parameter \(\delta\), we opt for the LAD estimator as the initial estimator, owing to its simplicity, distribution-free robustness, and computational stability.
	\end{rmk}
	The LAD estimator is asymptotically efficient under Laplace noise, while the LS estimator attains asymptotic efficiency under Gaussian noise. However, when the true noise distribution departs from these idealized forms, both estimators become suboptimal: their asymptotic variances exceed the CRLB. In such settings, the eMLE can substantially improve estimation accuracy by adaptively learning the noise distribution nonparametrically, thereby approaching the efficiency of the oracle MLE without requiring prior knowledge of the density function.
	
	To illustrate this, consider the noise following a Student’s \(t\)-distribution with degrees of freedom  \(s > 0\),  smaller values of \(s\) correspond to heavier tails. For the FIR system \eqref{fir}, the asymptotic covariance matrices of the LAD and LS estimators, as well as the CRLB  are given by
	\allowdisplaybreaks
	\begin{align*}
		&\Sigma_{LAD}^{-1}=\frac{s\pi \left[\Gamma\left(\frac{s}{2}\right)\right]^2}{4\sigma_u^2\left[\Gamma\left(\frac{s+1}{2}\right)\right]^2}I_p,~s>0,\\
		&\Sigma_{LS}^{-1}=\frac{s}{\sigma_u^2(s-2)}I_p, \,  ~s>2,\\
		&\Sigma_{CR}^{-1}=\frac{(s+3)}{\sigma_u^2(s+1)}I_p,~s>0,
	\end{align*}
	where $\Gamma(\cdot)$ is the Gamma function. 
	Several insights follow:
	\begin{enumerate}[(i)]
		\item The LAD estimator is well-defined and \(\sqrt{n}\)-consistent for all \(s > 0\), whereas the LS estimator only has a finite asymptotic variance when \(s > 2\).
		
		\item  LS outperforms LAD for \(s \geq 5\) , while LAD is more accurate for \(s < 5\).
		
		\item As \(s \to \infty\), the \(t\)-distribution approaches the Gaussian distribution and  the asymptotic variance of the LS estimator achieves the CRLB, confirming its efficiency in the Gaussian limit.
	\end{enumerate}
	The choice between the LAD and LS initial estimators is deferred to Section \ref{sec5.3}, where it is addressed jointly with the kernel bandwidth selection.
	
	\subsection{Joint selection of bandwidth and initial estimators}\label{sec5.3}
	The performance of the proposed eMLE and epMLE depends critically on the bandwidth $h$, analogous to nonparametric density estimation.
	For a fixed bandwidth \(h\), we denote the eMLE and epMLE by \(\widehat{\beta}^{eM}(h)\) and \(\widehat{\beta}^{epM}(h)\), respectively, and their efficient implementation using the one-step Newton-Raphson algorithms by \(\widehat{\beta}_{o}^{eM}(h)\) and \(\widehat{\beta}_{o}^{epM}(h)\).
	Meanwhile, the lack of prior knowledge regarding the noise distribution necessitates a data-driven choice of initial estimator. 
	The initial estimator plays a dual role: it is essential for the residual-based kernel density estimate in the eMLE, and it provides a stable starting point for the numerical optimization of the potentially non-convex objective function for both the eMLE and   epMLE.
	Therefore, this paper introduces a procedure to select both the bandwidth and the initial estimator, detailed for the eMLE and subsequently applied to the other estimators.

	For bandwidth selection, we propose a $K$-fold robust cross-validation method. Concretely, we first randomly partition the dataset into $K$ disjoint parts with equal or approximately equal sizes and denote the corresponding index sets by $\mathcal{I}_1,\ldots,\mathcal{I}_{K}$, respectively. Then we minimize the following  criterion function:
	\begin{align}\label{cvcrit}
		\text{CV}_{eM}(h)=\frac{1}{K}\sum_{j=1}^{K}\sum_{i\in\mathcal{I}_{j}}\rho\big(y_i-x_i^T\widehat{\beta}_{-\mathcal{I}_{j}}^{eM}(h)\big),
	\end{align}
	over the bandwidth $h$, where
	$\rho(\cdot)$ is a robust loss function such as the Huber loss function \eqref{huber},
	the eMLE estimate 
	\begin{align*}
		\widehat{\beta}_{-\mathcal{I}_{j}}^{eM}(h) =\mathop{\arg\min}_{\beta\in\Theta}\sum_{i\notin\mathcal{I}_{j}}^{n}\log\Bigg\{\frac{1}{nh}\sum_{k\notin\mathcal{I}_{j}}K\Big(\frac{\varepsilon_i(\beta)-\widehat{\varepsilon}_k}{h}\Big)\Bigg\}
	\end{align*} is obtained by leaving out the observations and the residuals $\widehat{\varepsilon}_k$ in $\mathcal{I}_{j}$. 
	Note that the bandwidth selection \eqref{cvcrit}  for the eMLE estimate depends on the choice of the initial estimator $\widehat{\beta}^{(0)}$ involved in the residuals $\widehat{\varepsilon}_k$.
	To reflect this dependence, we denote the bandwidth selection procedures by  $\text{CV}_{eM}^{LAD}$ and $\text{CV}_{eM}^{LS}$, corresponding to the use of the LAD and LS estimators as the initializers, respectively.
	Since the noise distribution may be heavy-tailed and have infinite variance, the ordinary quadratic loss function can lead to unstable bandwidth selection. We therefore adopt a robust loss function for the $K$-fold cross-validation procedure \cite{leung2005} to avoid such potential breakdown.
	
	Besides the bandwidth selection, we also need to determine an initial estimator between the LAD and LS due to unavailable noise distribution. 
	To make this joint selection, we introduce an adaptive cross-validation procedure, which proceeds as follows:
	\begin{enumerate}[(i)]
		\item For each candidate initializer (LS or LAD), compute the corresponding robust CV criterion by \eqref{cvcrit}.
		
		\item Select the pair (initializer, bandwidth) that yields the smallest CV value.
	\end{enumerate}
	This data-driven strategy adaptively learns from the observed samples, striking an effective balance between robustness and efficiency.

	\section{Simulations}\label{sec:6}
	To assess the performance of the proposed eMLE and epMLE estimators, we conduct extensive simulation studies,  comparing them with the LS  and LAD estimators. 
	We consider two simulation scenarios: Scenario A examines various noise types, including Gaussian, uniform, heavy-tailed, outlier-contaminated, and multimodal noise distributions; Scenario B investigates the tail behavior of noise using Student’s $t$-distributions with different degrees of freedom.

	\subsection{Data Bank}
	
	We first generate 100 FIR systems of order $p=5$, where all coefficients of each FIR are independently drawn from the uniform distribution $U[-2, 2]$.
	For each FIR system, we independently generate an input sequence of length $4000$ from the Gaussian distribution $\mathcal{N}(0,1)$
	and then produce the corresponding noise-free output sequence of length 4000 via convolution with the FIR system.
	This yields 100 noise-free output realizations.
	
	\subsubsection{Scenario A: different types of noise distributions}
	
	For each noise-free output, we generate four independent noise realizations of length 4000, drawn from following distributions:
	\begin{itemize}
		\item\textbf{Case 1: }\textbf{Gaussian:}
		$
		\varepsilon_i \sim \mathcal{N}(0, 1),
		$
		
		\item \textbf{Case 2: }\textbf{Uniform:}
		$
		\varepsilon_i \sim U[-2, 2],
		$
		
		\item \textbf{Case 3: }\textbf{Gaussian mixture (Outliers):}
		\\
		$\varepsilon_i \sim 0.9 \cdot \mathcal{N}(0, 1) + 0.1 \cdot \mathcal{N}(0, 100),
		$
		
		\item \textbf{Case 4: }\textbf{Gaussian mixture (Multimodal):}
		\\
		$\varepsilon_i \sim 0.5 \cdot \mathcal{N}(- 1.5, 1) + 0.5 \cdot \mathcal{N}(1.5, 1).
		$
	\end{itemize}
	Each noise sequence is then scaled to achieve a target signal-to-noise ratio (SNR), the ratio of the sample variance of the noise-free output sequence to that of the noise sequence.
	The target SNR is independently sampled from the uniform distribution $U[1,10]$ for each replication. 
	Finally, we add the scaled noise sequence to the corresponding noise-free output to obtain the noisy output sequence.
	
	\subsubsection{Scenario B: Student's t-distributions with various degrees of freedom}
	For each noise-free output, we  generate five independent noise sequences of length $4000$ from   Student's $t$-distributions with  degrees of freedom: $\nu=1,2,4,6,10$.
	To ensure comparability with Scenario A, we scale the noise sequences corresponding to  
	$\nu=4,6,10$  so that the resulting SNR matches that of Scenario A.
	In contrast, the noise sequences for  $\nu=1,2$ are not scaled as their theoretical variances are infinite.
	Finally, the noisy outputs are created by adding the resulting noise sequence to the noise-free output.

	\subsection{Implementation details and performance measure}
	We employ the Gaussian kernel 
	$K(t) = \frac{1}{\sqrt{2\pi}} \exp\left(-\frac{t^2}{2}\right)$  in our simulations. 
	We select the bandwidth by the 5-fold cross-validation using the Huber loss function with $\delta=1$. 
	
	We evaluate the performance of these methods by the squared estimation error (SEE)
	$$ {\rm SEE}(\widehat{\beta})=\|\widehat{\beta}-\beta_0\|^2,
	$$
	where $\widehat{\beta}$ is the resulting estimate and $\beta_0$ is the true value.  
	
	\subsection{Results on Scenario A}
	
	For Scenario A, we evaluate the following estimators:
	\begin{itemize}
		\item Classical: LS and  LAD.
		\item Proposed: eMLE, epMLE, oeMLE, and oepMLE, each implemented with three initialization strategies: LS-based, LAD-based, and adaptive initialization.
	\end{itemize}
	The average SEE for all methods, based on 100 Monte Carlo replications, is presented in Tables~\ref{table:1}–\ref{table:4} for sample sizes 
	$n=500,1000,2000,4000$ under four noise distributions.

	\begin{table}[!ht]
		\centering
		\caption{Case 1: $\mathcal{N}(0, 1)$}
		\label{table:1} 
			\begin{tabular}{cccccc|cc}
				\toprule
				\hline
				\multirow{3}{*}{init} & \multirow{3}{*}{$n$} & \multicolumn{6}{c}{Methods} \\\hline
				& & eMLE & oeMLE & epMLE & oepMLE & LAD & LS \\
				\hline
				{LAD} & \multirow{3}{*}{500} & 0.02035 & 0.02039 & 0.02130 & 0.02123 & \multirow{3}{*}{0.02938} & \multirow{3}{*}{0.01985} \\
				{LS} & & 0.02005 & 0.02007 & 0.02130 & 0.02084 & & \\
				{Adaptive} & & 0.02051 & 0.02056 & 0.02130 & 0.02157 & &\\
				\hline
				{LAD} & \multirow{3}{*}{1000} & 0.00982 & 0.00985 & 0.01011 & 0.01027 & \multirow{3}{*}{0.01527} & \multirow{3}{*}{0.00948} \\
				{LS} & & 0.00974 & 0.00976 & 0.01011 & 0.01010 & & \\
				{Adaptive} & & 0.00984 & 0.00981 & 0.01011 & 0.01040 & &\\
				\hline
				{LAD} & \multirow{3}{*}{2000} & 0.00549 & 0.00549 & 0.00546 & 0.00542 & \multirow{3}{*}{0.00892} & \multirow{3}{*}{0.00532} \\
				{LS} & & 0.00540 & 0.00540 & 0.00546 & 0.00543 & & \\
				{Adaptive} & & 0.00549 & 0.00553 & 0.00546 & 0.00544 & &\\
				\hline
				{LAD} & \multirow{3}{*}{4000} & 0.00261 & 0.00262 & 0.00261 & 0.00261 & \multirow{3}{*}{0.00489} & \multirow{3}{*}{0.00251} \\
				{LS} & & 0.00258 & 0.00258 & 0.00261 & 0.00261 & & \\
				{Adaptive} & & 0.00261 & 0.00261 & 0.00261 & 0.00261 & &\\
				\hline
				\bottomrule
			\end{tabular}%

	\end{table}
	
	\begin{table}[!ht]
		\centering
		\caption{Case 2: $U[-2, 2]$}
		\label{table:2} 
			\begin{tabular}{cccccc|cc}
				\toprule
				\hline
				\multirow{3}{*}{init} & \multirow{3}{*}{$n$} & \multicolumn{6}{c}{Methods} \\\hline
				& & eMLE & oeMLE & epMLE & oepMLE & LAD & LS \\
				\hline
				{LAD} & \multirow{3}{*}{500} & 0.00770 & 0.00769 & 0.00740 & 0.00777 & \multirow{3}{*}{0.04436} & \multirow{3}{*}{0.01418} \\
				{LS} & & 0.00710 & 0.00698 & 0.00740 & 0.00776 & & \\
				{Adaptive} & & 0.00745 & 0.00725 & 0.00742 & 0.00738 & &\\
				\hline
				{LAD} & \multirow{3}{*}{1000} & 0.00364 & 0.00373 & 0.00358 & 0.00365 & \multirow{3}{*}{0.02219} & \multirow{3}{*}{0.00692} \\
				{LS} & & 0.00351 & 0.00354 & 0.00358 & 0.00371 & & \\
				{Adaptive} & & 0.00357 & 0.00362 & 0.00358 & 0.00340 & &\\
				\hline
				{LAD} & \multirow{3}{*}{2000} & 0.00186 & 0.00186 & 0.00193 & 0.00190 & \multirow{3}{*}{0.01108} & \multirow{3}{*}{0.00359} \\
				{LS} & & 0.00183 & 0.00183 & 0.00193 & 0.00192 & & \\
				{Adaptive} & & 0.00184 & 0.00184 & 0.00193 & 0.00191 & &\\
				\hline
				{LAD} & \multirow{3}{*}{4000} & 0.00097 & 0.00097 & 0.00101 & 0.00100 & \multirow{3}{*}{0.00506} & \multirow{3}{*}{0.00196} \\
				{LS} & & 0.00096 & 0.00096 & 0.00101 & 0.00102 & & \\
				{Adaptive} & & 0.00097 & 0.00097 & 0.00101 & 0.00100 & &\\
				\hline
				\bottomrule
			\end{tabular}%
 
	\end{table}
	
	\begin{table}[!ht]
		\centering
		\caption{Case 3: $0.9 \cdot \mathcal{N}(0, 1) + 0.1 \cdot \mathcal{N}(0, 100)$}
		\label{table:3} 
			\begin{tabular}{cccccc|cc}
				\toprule
				\hline
				\multirow{3}{*}{init} & \multirow{3}{*}{$n$} & \multicolumn{6}{c}{Methods} \\\hline
				& & eMLE & oeMLE & epMLE & oepMLE & LAD & LS \\
				\hline
				{LAD} & \multirow{3}{*}{500} & 0.00179 & 0.00180 & 0.00159 & 0.00160 & \multirow{3}{*}{0.00225} & \multirow{3}{*}{0.01322} \\
				{LS} & & 0.00179 & 0.00180 & 0.00159 & 0.00173 & & \\
				{Adaptive} & & 0.00173 & 0.00173 & 0.00159 & 0.00160 & &\\
				\hline
				{LAD} & \multirow{3}{*}{1000} & 0.00093 & 0.00093 & 0.00083 & 0.00083 & \multirow{3}{*}{0.00113} & \multirow{3}{*}{0.00764} \\
				{LS} & & 0.00095 & 0.00095 & 0.00083 & 0.00087 & & \\
				{Adaptive} & & 0.00094 & 0.00094 & 0.00083 & 0.00087 & &\\
				\hline
				{LAD} & \multirow{3}{*}{2000} & 0.00047 & 0.00047 & 0.00041 & 0.00041 & \multirow{3}{*}{0.00059} & \multirow{3}{*}{0.00335} \\
				{LS} & & 0.00047 & 0.00047 & 0.00041 & 0.00042 & & \\
				{Adaptive} & & 0.00046 & 0.00046 & 0.00041 & 0.00041 & &\\
				\hline
				{LAD} & \multirow{3}{*}{4000} & 0.00026 & 0.00026 & 0.00023 & 0.00023 & \multirow{3}{*}{0.00031} & \multirow{3}{*}{0.00180} \\
				{LS} & & 0.00026 & 0.00026 & 0.00023 & 0.00023 & & \\
				{Adaptive} & & 0.00026 & 0.00026 & 0.00023 & 0.00023 & &\\
				\hline
				\bottomrule
			\end{tabular}%
	\end{table}
	
	\begin{table}[!ht]
		\centering
		\caption{Case 4: $0.5 \cdot \mathcal{N}(-1.5, 1) + 0.5 \cdot \mathcal{N}(1.5, 1)$}
		\label{table:4} 
			\begin{tabular}{cccccc|cc}
				\toprule
				\hline
				\multirow{3}{*}{init} & \multirow{3}{*}{$n$} & \multicolumn{6}{c}{Methods} \\\hline
				& & eMLE & oeMLE & epMLE & oepMLE & LAD & LS \\
				\hline
				{LAD} & \multirow{3}{*}{500} & 0.01068 & 0.01062 & 0.00983 & 0.01075 & \multirow{3}{*}{0.05038} & \multirow{3}{*}{0.01396} \\
				{LS} & & 0.01016 & 0.01013 & 0.00983 & 0.01023 & & \\
				{Adaptive} & & 0.01046 & 0.01020 & 0.00983 & 0.01027 & &\\
				\hline
				{LAD} & \multirow{3}{*}{1000} & 0.00495 & 0.00497 & 0.00484 & 0.00493 & \multirow{3}{*}{0.02698} & \multirow{3}{*}{0.00680} \\
				{LS} & & 0.00495 & 0.00497 & 0.00484 & 0.00492 & & \\
				{Adaptive} & & 0.00504 & 0.00503 & 0.00484 & 0.00491 & &\\
				\hline
				{LAD} & \multirow{3}{*}{2000} & 0.00224 & 0.00222 & 0.00216 & 0.00216 & \multirow{3}{*}{0.01399} & \multirow{3}{*}{0.00316} \\
				{LS} & & 0.00221 & 0.00220 & 0.00216 & 0.00218 & & \\
				{Adaptive} & & 0.00223 & 0.00222 & 0.00216 & 0.00216 & &\\
				\hline
				{LAD} & \multirow{3}{*}{4000} & 0.00113 & 0.00112 & 0.00111 & 0.00108 & \multirow{3}{*}{0.00725} & \multirow{3}{*}{0.00163} \\
				{LS} & & 0.00112 & 0.00112 & 0.00111 & 0.00112 & & \\
				{Adaptive} & & 0.00112 & 0.00112 & 0.00111 & 0.00110 & &\\
				\hline
				\bottomrule
			\end{tabular}%
	\end{table}
	
	Under Gaussian noise, LS performs best as expected, and the proposed methods are significantly better than LAD and comparable to the optimal LS. For the remaining three noise types, the proposed methods consistently outperform both LS and LAD across all sample sizes, and the relative improvement over the better of the two baseline methods is substantial:
	\begin{itemize} 
		\item \textbf{Case 2:}  the improvement over LS ranges from 45.2\% to 50.8\%;
		
		\item \textbf{Case 3:}  the improvement over   LAD ranges from 16.5\% to 30.8\%;
		
		\item \textbf{Case 4:}  the improvement over LS ranges from 23.0\% to 33.4\%.
	\end{itemize}
	These gains highlight the adaptability and robustness of the proposed framework under non-Gaussian, heavy-tailed, or structured noise.

	\subsection{Results on Scenario B}
	
	For Scenario B, we evaluate the following estimators:
	\begin{itemize}
		\item Classical:  LS, LAD, and the tMLE (the MLE with true density function initialized at the true parameters);
		\item Proposed: eMLE, epMLE, oeMLE, and oepMLE, each implemented with  adaptive initialization.
	\end{itemize}
	
\begin{figure}[htbp]
		\centering
		\includegraphics[width=0.85\textwidth]{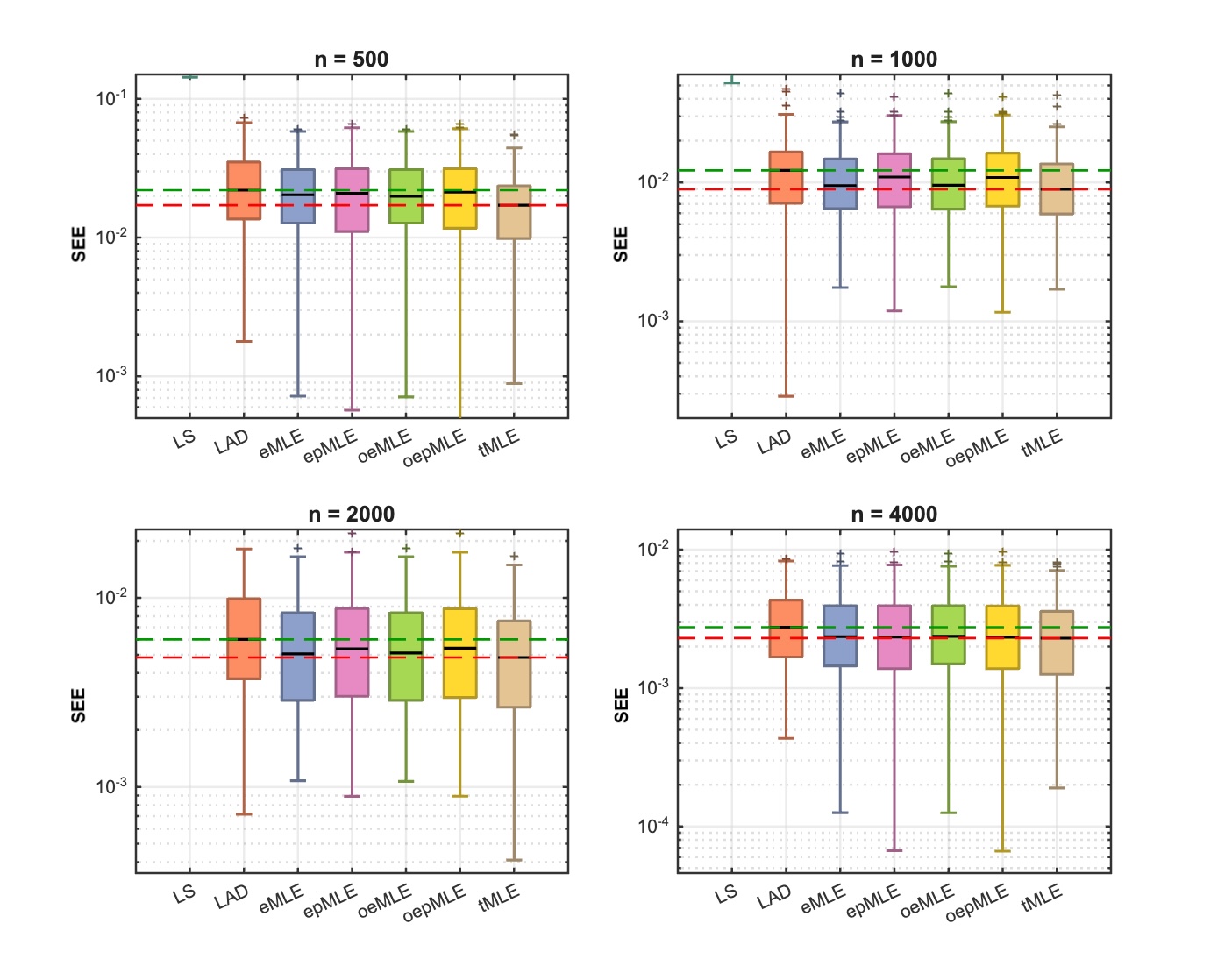}  
		\caption{Boxplots of SEE under $t(1)$-noise.}
		\label{fig:1}
	\end{figure}

\begin{figure}[htbp]
		\centering
		\includegraphics[width=0.85\textwidth]{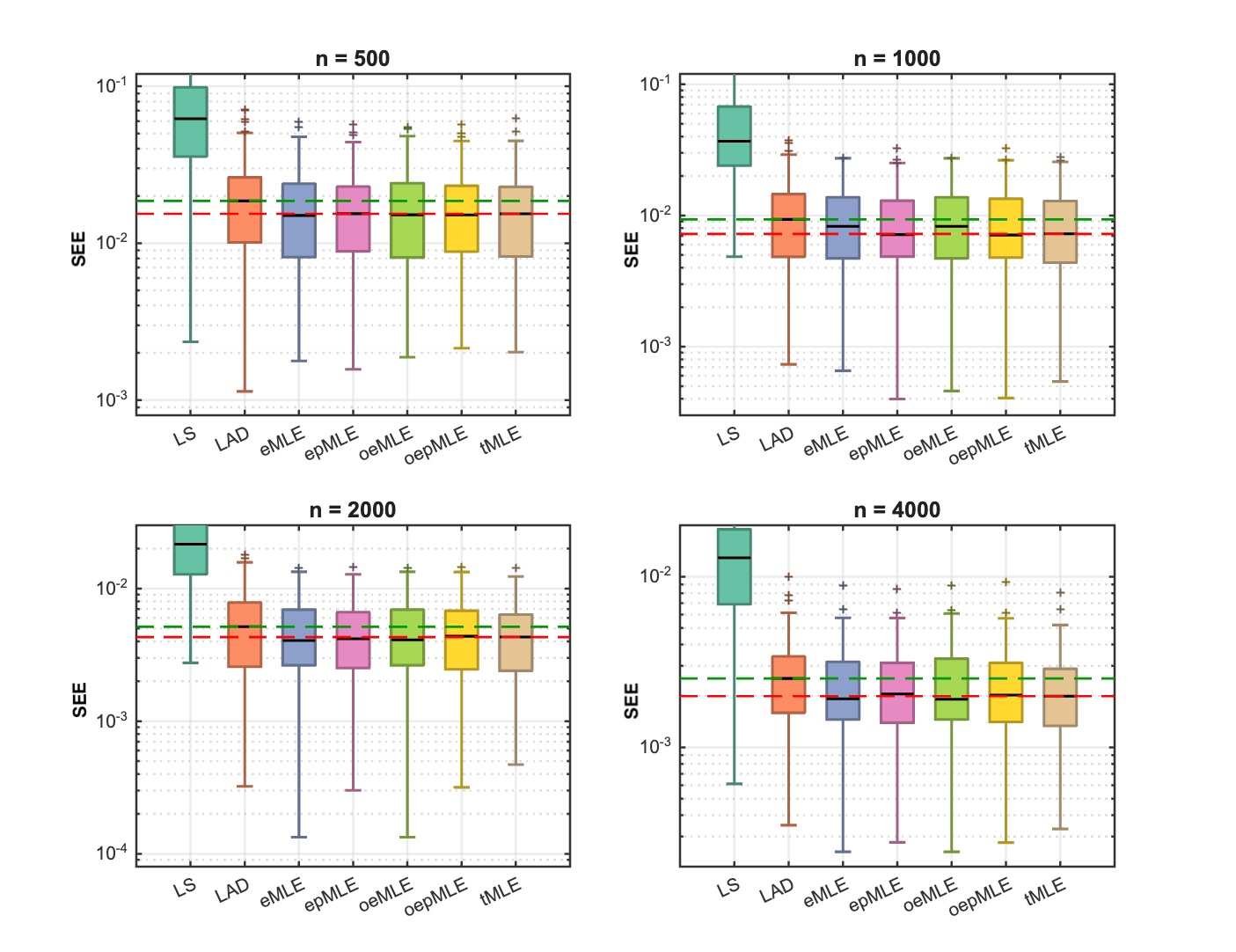}  
		\caption{Boxplots of SEE under $t(2)$-noise.}
		\label{fig:2}
	\end{figure}

\begin{figure}[htbp]
		\centering
		\includegraphics[width=0.85\textwidth]{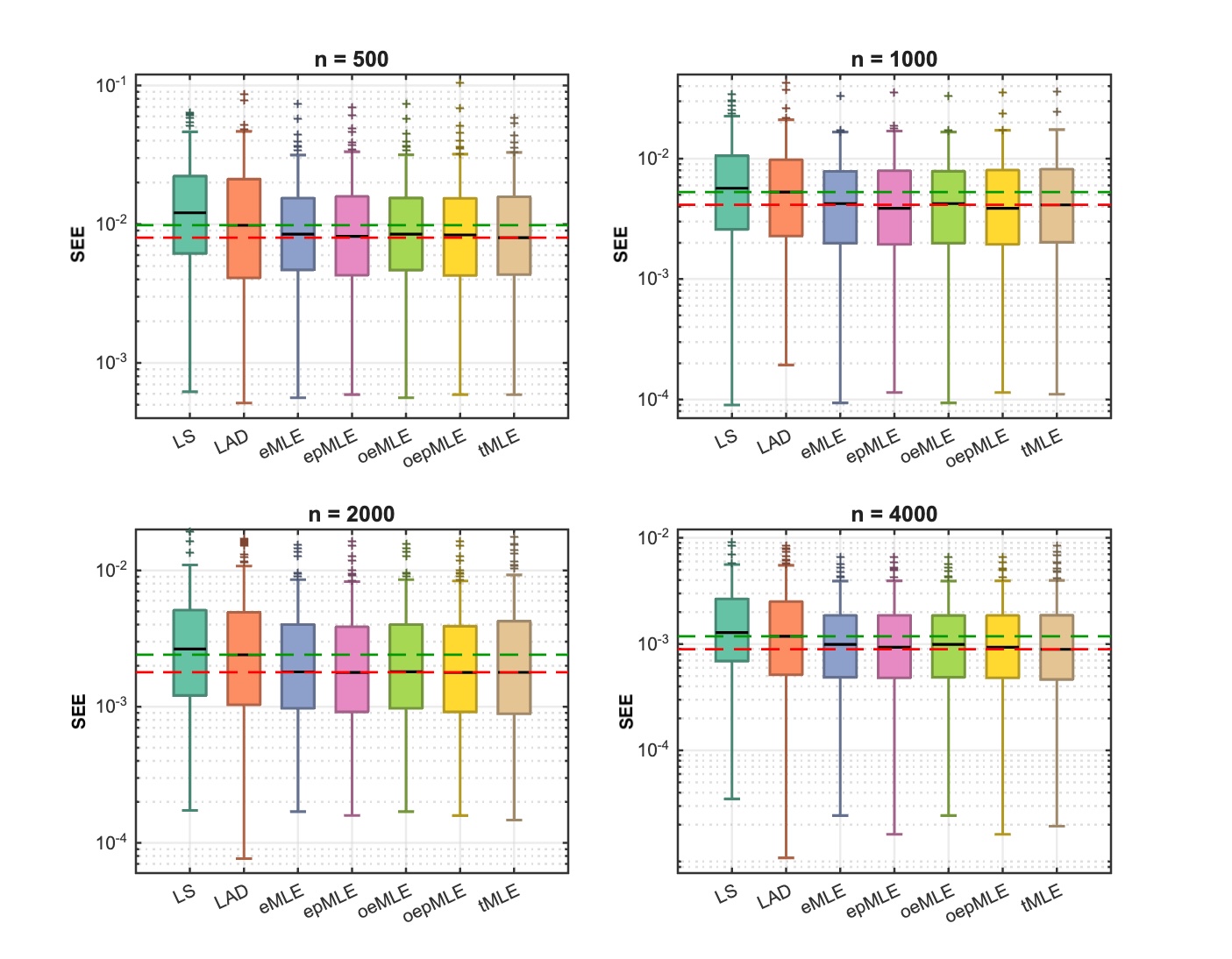}  
		\caption{Boxplots of SEE under $t(4)$-noise.}
		\label{fig:3}
	\end{figure}

\begin{figure}[htbp]
		\centering
		\includegraphics[width=0.85\textwidth]{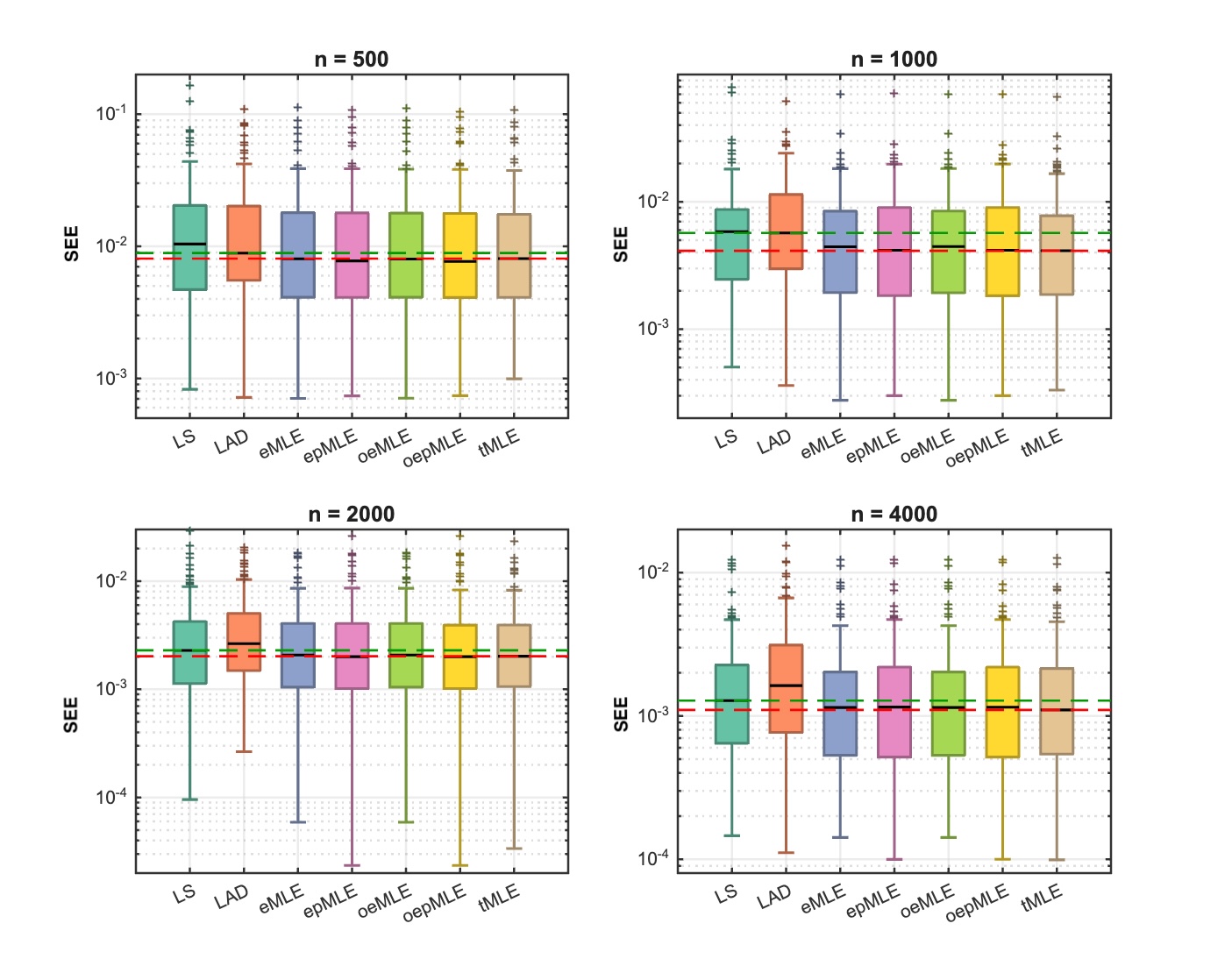}  
		\caption{Boxplots of SEE under $t(6)$-noise.}
		\label{fig:4}
	\end{figure}

\begin{figure}[htbp]
		\centering
		\includegraphics[width=0.85\textwidth]{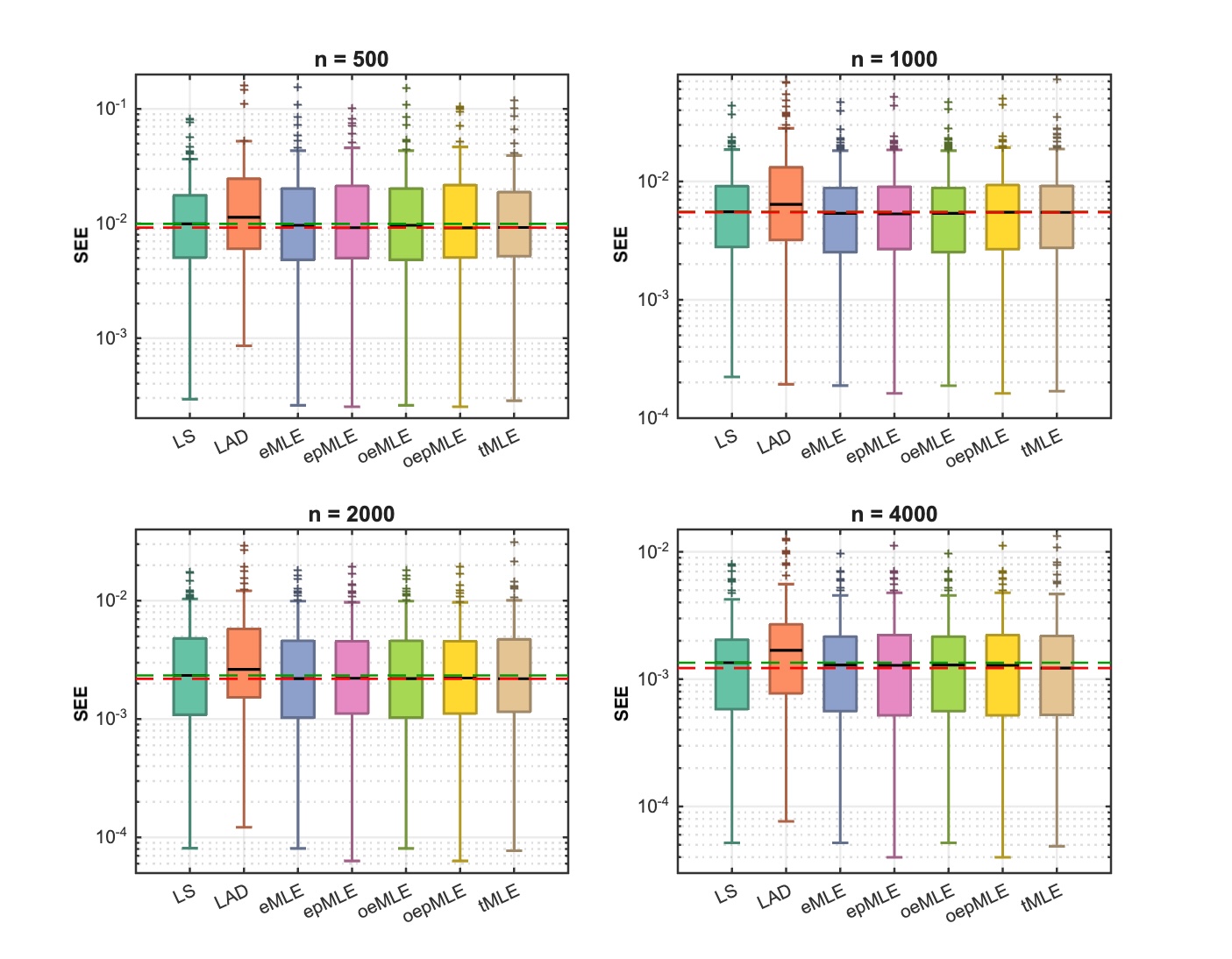}  
		\caption{Boxplots of SEE under $t(10)$-noise.}
		\label{fig:5}
	\end{figure}

	Figures~\ref{fig:1}–\ref{fig:5} display boxplots of the SEE for the considered estimators across 100 Monte Carlo replications, with sample  sizes $n=500,1000,2000,4000$
	and degrees of freedom $\nu=1,2,4,6,10$.
	To facilitate comparison, dotted lines indicate key benchmarks: the median of the LAD estimator (for $\nu=1,2,4$) or the LS estimator (for $\nu=6,10$) is shown in green, and the median of the tMLE is shown in red.

	All four proposed estimators with adaptive initialization (eMLE, epMLE, oeMLE, oepMLE) uniformly outperform both LS and LAD  across all tested sample sizes and degrees of freedom.
	Their performance is nearly comparable to the asymptotically optimal tMLE for $\nu=2,4,6,10$, particularly in large-sample settings. Notably, even in the extreme heavy-tailed case of Cauchy noise ($\nu=1$), where the absolute mean is infinite, our estimators  closely approach tMLE at
	$n=4000$, confirming their asymptotic efficiency.

	\section{Conclusion}
	\label{sec:7}
	In this paper, we have presented eMLE and epMLE, two asymptotically efficient estimators for FIR system identification under unknown noise distributions. By integrating nonparametric kernel density estimation into the likelihood framework, both methods achieve consistency and asymptotic efficiency without imposing parametric assumptions and moment conditions on the noise. To enhance computational tractability, we introduce one-step Newton--Raphson algorithms and provide practical guidelines for joint bandwidth selection and  initialization. Extensive simulations across heavy-tailed, multimodal, and outlier-contaminated noise settings demonstrate that the proposed estimators consistently outperform classical LS and LAD, while remaining competitive with oracle MLE with known noise density function. Theoretically grounded and empirically validated, our approach offers a statistically efficient and robust solution for complex noise distributions.

	\section*{Appendix}
	
	The Appendix is divided into four parts: Appendix A contains the main threads of the proofs of theorems, Appendix B includes several lemmas used in proofs of theorems, Appendix C provides results on the uniform convergence of the estimated KDE, and Appendix D presents some useful auxiliary lemmas.
	
	\section*{Appendix A: Proofs of  main results}
	\renewcommand{\theequation}{A\arabic{equation}}
	\setcounter{equation}{0}
	
	\setcounter{thm}{0}
	\renewcommand{\theorem}{A\arabic{thm}}
	\setcounter{lem}{0}
	\renewcommand{\theorem}{A\arabic{lem}}
	\setcounter{rmk}{0}
	\renewcommand{\theorem}{A\arabic{rmk}}
	\renewcommand{\thesection}{A}
	\def\thesectiondis{\thesection.}                   
	\def\thesubsectiondis{\thesection.\arabic{subsection}.}          
	\def\thesubsubsectiondis{\thesubsection.\arabic{subsubsection}.}
	\setcounter{subsection}{0}
	
	\subsection{Proof of Theorem \ref{thm1}}
	For notational consistency, we denote the log-likelihood function with true noise density by
	$$  
	L_n^{eM}(\beta)\eq\frac{1}{n}\sum_{k=1}^{n}\log f_o(y_k-x_k^T\beta).
	$$
	
	\textbf{1. Prove $\widehat{\beta}^{eM}$ converges to $\beta_0$ in probability.}
	
	We first prove that
	$
	\sup\limits_{\beta\in \Theta}\left|\widehat{L}_n^{eM}(\beta)-L_{eM}(\beta)\right|=o_p(1)
	$.
	
	On one hand, by Assumptions \ref{asp2}, \ref{asp3}-(\romannumeral2), Lemma \ref{lem:c2} yields 
	$$
	\sup_{\beta\in\Theta}\left|L_n^{eM}(\beta)-L_{eM}(\beta)\right|=o_p(1).$$
	
	On the other hand, by Assumption \ref{asp2}-(\romannumeral3) and Lemma \ref{lem:c0}, 
	\begingroup
	\allowdisplaybreaks
	\begin{align*}
		\sup_{\beta\in \Theta}\Big|
		\widehat{L}_n^{eM}(\beta)-
		L_n^{eM}(\beta)\Big|&=\sup_{\beta\in \Theta}\Bigg|\frac{1}{n}\sum_{k=1}^{n}\log\frac{\widehat{f}_h(\varepsilon_k(\beta))}{f_o(\varepsilon_k(\beta))}\Bigg| \leq\sup_{t\in\mathcal{T}}\Bigg|\log\Big(1+\frac{\widehat{f}_h(t)-f_o(t)}{f_o(t)}\Big)\Bigg|\\
		&\leq\frac{\sup\limits_{t\in\mathcal{T}}|\widehat{f}_h(t)-f_o(t)|}{\inf\limits_{t\in\mathcal{T}}f_o(t)-\sup\limits_{t\in\mathcal{T}}|\widehat{f}_h(t)-f_o(t)|},
	\end{align*}
	\endgroup
	which is $o_p(1)$ in terms of Lemma \ref{lem:b1}.
	
	Moreover, by Assumption \ref{asp3}-(\romannumeral1) and the definition of $\widehat{\beta}^{eM}$, Lemma \ref{lem:c1} yields that $\widehat{\beta}^{eM}\xrightarrow[]{p}\beta_0$.
	
	\textbf{2. Prove the asymptotic normality of $\widehat{\beta}^{eM}$}. 
	
	By  the Taylor expansion of $\widehat{S}_{eM}(\beta)$ given in \eqref{sem} around $\beta_0$, we have
	\begin{align}\label{barbeta_em}
		\widehat{S}_{eM}(\widehat{\beta}^{eM})=\widehat{S}_{eM}(\beta_0)+\widehat{H}_{eM}(\beta^*)(\widehat{\beta}^{eM}-\beta_0),
	\end{align}
	where $\beta^{*}$ lies between $\widehat{\beta}^{eM}$ and $\beta_0$.
	Meanwhile, the maximizer $\widehat{\beta}^{eM}$  of  $\widehat{L}_n^{eM}(\beta)$ satisfies
	$
	\widehat{S}_{eM}(\widehat{\beta}^{eM})=0.
	$
	
	By Lemmas \ref{lem:d1} and \ref{lem:d2}, as $n\xra{}\infty$, we have 
	$$
	\sqrt{n}\widehat{S}_{eM}(\beta_0)\xrightarrow[]{d}\mathcal{N}(0,\Sigma),
	~~\widehat{H}_{eM}(\beta^*)\xrightarrow{p}\Sigma.
	$$
	Therefore, Lemma \ref{lem:c3} implies that
	$
	\sqrt{n}(\widehat{\beta}^{eM}-\beta_0)\xrightarrow[]{d}\mathcal{N}(0,\Sigma^{-1}).
	$

	\subsection{Proof of Theorem \ref{thm2}}
	
	Similarly,  for the proof of the epMLE estimator, we denote
	\begin{align}
		\nonumber
		&{L}_n^{epM}(\beta)\eq\frac{1}{n}\sum_{k=1}^{n}\log{f}(y_k-x_k^T\beta,\beta),\\
		\label{ftbeta}
		&f(t,\beta)\eq\int f_o(t-x^T(\beta_0-\beta))\d F(x),
	\end{align}
	where $F(\cdot)$ is the joint distribution function of the regressor $x_{k}=[u_{k-1},\ldots,u_{k-p}]^T$.
	
	\textbf{1. Prove the consistency of the epMLE estimator $\widehat{\beta}^{epM}$}.
	
	By Lemma \ref{lem:b2}, we have
	$$
	\sup_{t\in\mathcal{T},~\beta\in\Theta}|\widehat{f}_h(t,\beta)-f(t,\beta)|=o_p(1).
	$$
	By Assumption \ref{asp2}-(\romannumeral3),  we have $\inf_{t\in\mathcal{T},~\beta\in\Theta}f(t,\beta)>0$, and further by Lemma \ref{lem:c0}, there holds that
	\begin{align*}
		\sup_{\beta\in\Theta}\left|\widehat{L}_{n}^{epM}(\beta)-L_n^{epM}(\beta)\right| &\leq\sup_{t\in\mathcal{T},\beta\in\Theta}\Bigg|\log\Big(1+\frac{\widehat{f}_h(t,\beta)-f(t,\beta)}{f(t,\beta)}\Big)\Bigg|\\
		&\leq\frac{\sup\limits_{t\in\mathcal{T},\beta\in\Theta}|\widehat{f}_h(t,\beta)-f(t,\beta)|}{\inf\limits_{t\in\mathcal{T},\beta\in\Theta} f(t,\beta)-\sup\limits_{t\in\mathcal{T},\beta\in\Theta}|\widehat{f}_h(t,\beta)-f(t,\beta)|}=o_p(1).
	\end{align*}
	
	By Assumptions \ref{asp2}, \ref{asp7}-(\romannumeral2) and Lemma \ref{lem:c2}, we have
	$$
	\sup_{\beta\in\Theta}|L_{n}^{epM}(\beta)-L_{epM}(\beta)|=o_{p}(1).
	$$
	
	Combining the above results and by the triangle inequality, we obtain
	$$
	\sup_{\beta\in\Theta}\left|\widehat{L}_{n}^{epM}(\beta)-L_{epM}(\beta)\right|=o_p(1).
	$$
	Finally, by Assumption \ref{asp7}-(\romannumeral1) and Lemma \ref{lem:c1}, we establish the consistency, i.e., $\widehat{\beta}^{epM}\xrightarrow{p}\beta_0$. 
	
	\textbf{2. Prove the asymptotic normality of $\widehat{\beta}^{epM}$}.
	
	Consider the Taylor expansion of $\widehat{S}_{epM}(\beta)$ given in \eqref{sepm} at $\beta_0$
	\begin{align}\label{barbeta_epm}
		\widehat{S}_{epM}(\widehat{\beta}^{epM})=\widehat{S}_{epM}(\beta_0)+\widehat{H}_{epM}(\bar{\beta})(\widehat{\beta}^{epM}-\beta_0),
	\end{align}
	where $\bar{\beta}$ lies between $\widehat{\beta}^{epM}$ and $\beta_0$.
	Meanwhile, the maximizer $\widehat{\beta}^{epM}$  of  $\widehat{L}_{epM}(\beta)$ satisfies
	$
	\widehat{S}_{epM}(\widehat{\beta}^{epM})=0.
	$
	Firstly, by Lemma 1 in \cite{yuan2007} and $\bar{\beta}\xrightarrow[]{p}\beta_0$,
	we have
	$
	\widehat{H}_{epM}(\bar{\beta})\xrightarrow[]{p}\Sigma.
	$
	Secondly,  Lemma \ref{lem:d3} indicates
	$
	\sqrt{n}\widehat{S}_{epM}(\beta_0)\xrightarrow[]{d}\mathcal{N}(0,\Sigma).
	$
	Finally,  Lemma \ref{lem:d2} entails 
	$$
	\sqrt{n}(\widehat{\beta}^{epM}-\beta_0)\xrightarrow{d}\mathcal{N}(0,\Sigma^{-1}).
	$$

	\subsection{Proof of Theorem \ref{thm3}}
	
	By the Taylor expansion of $\widehat{S}_{eM}(\beta)$ around $\widehat{\beta}^{(0)}$
	\begin{align*}
		- \widehat{S}_{eM}(\widehat{\beta}^{(0)}) 
	 =
		\widehat{S}_{eM}(\widehat{\beta}^{eM}) - \widehat{S}_{eM}(\widehat{\beta}^{(0)}) 
		= \widehat{H}_{eM}(\widehat{\beta}^{(0)}) \big( \widehat{\beta}^{eM} \!-\! \widehat{\beta}^{(0)} \big)+ o_{p}\big( \|\widehat{\beta}^{eM} \!-\!\widehat{\beta}^{(0)}\| \big).
	\end{align*}
	For large $n$, by Assumption \ref{asp3}, we have $\widehat{H}_{eM}(\widehat{\beta}^{(0)})$ is non-singular, and hence there holds that
	\begin{align*}
		&\sqrt{n}(\widehat{\beta}_o^{eM} -\widehat{\beta}^{eM})=o_{p}\left( \sqrt{n}\|\widehat{\beta}^{eM} - \widehat{\beta}^{(0)}\| \right).
	\end{align*}
	By triangle inequality, we have
	$$
	\|\widehat{\beta}^{eM}-\widehat{\beta}^{(0)}\|\leq\|\widehat{\beta}^{eM}-\beta_0\|+\|\beta_0-\widehat{\beta}^{(0)}\|=O_{p}(n^{-1/2}).
	$$
	This implies that
	$\sqrt{n}(\widehat{\beta}_o^{eM} -\widehat{\beta}^{eM})=o_p(1)$.
	Therefore, we have proved that $
	\sqrt{n}(\widehat{\beta}_{o}^{eM}-\beta_0)\xrightarrow[]{d}\mathcal{N}(0,\Sigma^{-1})
	$ in terms of the asymptotic normality of $\widehat{\beta}^{eM}$ presented in Theorem \ref{thm1}.

	\subsection{Proof of Theorem \ref{thm4}}
	The proof follows similar arguments in the proof of Theorem \ref{thm3} and is therefore omitted.
	
	\renewcommand{\theequation}{B\arabic{equation}}
	\setcounter{equation}{0}
	\setcounter{thm}{0}
	\renewcommand{\thethm}{B\arabic{thm}}
	\setcounter{lem}{0}
	\renewcommand{\thelem}{B\arabic{lem}}
	\setcounter{rmk}{0}
	\renewcommand{\thermk}{B\arabic{rem}}
	\setcounter{subsection}{0}
	\renewcommand{\thesection}{B}
	\section*{Appendix B: Lemmas used in the proofs of theorems}
	
	\begin{lem}\label{lem:d1}
		Under Assumptions in Theorem \ref{thm1}, we have
		$$
		\sqrt{n}\widehat{S}_{eM}(\beta_0)\xrightarrow[]{d}\mathcal{N}(0,\Sigma).
		$$
	\end{lem}
	\begin{proof}
		By equation \eqref{hat-true-2} in the Appendix D for $g=\widehat{f}_h$, we have
		\begingroup
		\allowdisplaybreaks
		\begin{align*}
			&\widehat{S}_{eM}(\beta_0)=\frac{1}{n}\sum_{k=1}^{n}\frac{\widehat{f}_h^{(1)}(\varepsilon_k)}{\widehat{f}_h(\varepsilon_k)}x_k\\
			&=\frac{1}{n}\sum_{k=1}^{n}\widehat{f}_h^{(1)}(\varepsilon_k)\Big(\frac{1}{f_o(\varepsilon_k)}-\frac{\widehat{f}_h(\varepsilon_k)-f_o(\varepsilon_k)}{f_o^2(\varepsilon_k)} +\frac{(\widehat{f}_h(\varepsilon_k)-f_o(\varepsilon_k))^2}{f_o^2(\varepsilon_k)\widehat{f}_h(\varepsilon_k)}\Big)x_k\\
			&=\frac{1}{n}\sum_{k=1}^{n}\frac{f_o^{(1)}(\varepsilon_k)}{f_o(\varepsilon_k)}x_k+\frac{1}{n}\sum_{k=1}^{n}\frac{\widehat{f}_h^{(1)}(\varepsilon_k)-f_o^{(1)}(\varepsilon_k)}{f_o(\varepsilon_k)}x_k\\
			&\hspace{5mm}-\frac{1}{n}\sum_{k=1}^{n}\frac{\widehat{f}_h^{(1)}(\varepsilon_k)(\widehat{f}_h(\varepsilon_k)-f_o(\varepsilon_k))}{f_o^2(\varepsilon_k)}x_k +\frac{1}{n}\sum_{k=1}^{n}\frac{\widehat{f}_h^{(1)}(\varepsilon_k)(\widehat{f}_h(\varepsilon_k)-f_o(\varepsilon_k))^2}{f_o^2(\varepsilon_k)\widehat{f}_h(\varepsilon_k)}x_k\\
			&\hspace{-0mm}\eq I_1+I_2-I_3+I_4.
		\end{align*}
		\endgroup
		
		For $I_1$, $\{\big( f_o^{(1)}(\varepsilon_k)\big/f_o(\varepsilon_k)\big)x_k\}_{k=1}^{n}$ is a $p$-dependent sequence with zero mean and bounded variance.
		Thus, we have
		$
		\sqrt{n}I_1\xrightarrow[]{d}\mathcal{N}(0,\Sigma)
		$
		by central limit theorem for the $p$-dependent sequence \cite[Theorem 9.1]{dasgupta2008}. In the following, we bound terms $I_2,I_3,I_4$ to be $o_{p}(n^{-1/2})$.
		
		For $I_2$, we have the decomposition
		\allowdisplaybreaks
		\begin{align*}
			I_2 =\frac{1}{n}\sum_{k=1}^{n}\frac{\widehat{f}_h^{(1)}(\varepsilon_k)-\widetilde{f}_h^{(1)}(\varepsilon_k)}{f_o(\varepsilon_k)}x_k +\frac{1}{n}\sum_{k=1}^{n}\frac{\widetilde{f}_h^{(1)}(\varepsilon_k)-f_o^{(1)}(\varepsilon_k)}{f_o(\varepsilon_k)}x_k\eq I_{21}+I_{22},
		\end{align*}
		where
		\begin{align}\label{kde'true}
			\widetilde{f}_h^{(1)}(t)\eq\frac{1}{nh^{2}}\sum_{k=1}^{n}K^{(1)}\Big(\frac{t-\varepsilon_k}{h}\Big),
		\end{align}

		Firstly, the term $I_{21}$ is decomposed as by Taylor expansion
		\allowdisplaybreaks
		\begin{align*}
			I_{21}&=\frac{1}{n^2h^2}\sum_{k,j=1}^{n}
			\frac{\left\{
				K^{(1)} \big(\frac{\varepsilon_k-\widehat{\varepsilon}_j}{h}\big)
				-K^{(1)}\big(\frac{\varepsilon_k-\varepsilon_j}{h}\big)\right\}}{f_o(\varepsilon_k)}x_k\\
			&=\frac{1}{n^2h^2}\sum_{k,j=1}^{n}\frac{K^{(2)}\big(\frac{\varepsilon_k-\varepsilon_j}{h}\big)}{f_o(\varepsilon_k)}\Big(\frac{x_j^T(\widehat{\beta}^{(0)}-\beta_0)}{h}\Big)x_k\\
			&\hspace{4mm}+\frac{1}{2n^2h^2}\sum_{k,j=1}^{n}\frac{K^{(3)}\big(\frac{\varepsilon_k-\varepsilon_j}{h}\big)}{f_o(\varepsilon_k)}\Big(\frac{x_j^T(\widehat{\beta}^{(0)}-\beta_0)}{h}\Big)^2x_k\\
			&\hspace{4mm}+\frac{1}{6n^2h^2}\sum_{k,j=1}^{n}\frac{K^{(4)}\big(\frac{\varepsilon_k-\varepsilon_j^*}{h}\big)}{f_o(\varepsilon_k)}\Big(\frac{x_j^T(\widehat{\beta}^{(0)}-\beta_0)}{h}\Big)^3x_k\\
			&\hspace{-2mm}\eq I_{211}+I_{212}+I_{213},
		\end{align*}
		where $\varepsilon_j^*$ is a point between $\widehat{\varepsilon}_j$ and $\varepsilon_j$.
		
		As $p$ is finite, it suffices to bound each entry of $I_{211}$
		$$
		\frac{1}{n^2h^2}\sum_{k,j=1}^{n}\frac{K^{(2)}\big(\frac{\varepsilon_k-\varepsilon_j}{h}\big)}{f_o(\varepsilon_k)}\Big(\frac{x_j^T(\widehat{\beta}^{(0)}-\beta_0)}{h}\Big)u_{k-l},
		$$
		for $l=1,\cdots,p$,
		which is rewritten as
		$$
		\sum_{i=1}^{p}(\widehat{\beta}^{(0)}-\beta_0)_{i}
		\bigg(\frac{1}{n^2h^3}\sum_{k,j=1}^{n}\frac{K^{(2)}\big(\frac{\varepsilon_k-\varepsilon_j}{h}\big)}{f_o(\varepsilon_k)}u_{k-l}u_{j-i}\bigg),
		$$
		where $(\widehat{\beta}^{(0)}-\beta_0)_{i}$ denotes the $i$-th entry of $\widehat{\beta}^{(0)}-\beta_0$.
		Let us consider
		$$
		I_{211}'\eq\frac{1}{n^2h^3}\sum_{k,j=1}^{n}\frac{K^{(2)}\big(\frac{\varepsilon_k-\varepsilon_j}{h}\big)}{f_o(\varepsilon_k)}u_{k-l}u_{j-i}
		$$
		for $l,i=1,\cdots,p$. 
		Thus, we have that $\mathbb{E}[I_{211}'^2]$ is
		\begin{align*}
			\frac{1}{n^4h^6}\sum_{k_1,k_2,j_1,j_2=1}^{n}
			&\mathbb{E}\bigg[\frac{K^{(2)}\big(\frac{\varepsilon_{k_1}-\varepsilon_{j_1}}{h}\big)
				K^{(2)}\big(\frac{\varepsilon_{k_2}-\varepsilon_{j_2}}{h}\big)}{f_o(\varepsilon_{k_1})f_o(\varepsilon_{k_2})}\bigg]\\
			&\times\mathbb{E}[u_{k_1-l}u_{k_2-l}u_{j_1-i}u_{j_2-i}].
		\end{align*}
		The expectation of the terms associated with the input is non-zero if and only if the indices $k_1-l,k_2-l,j_1-i,j_2-i$ are all the same or there are two pairs of indices, the number of these terms is at most $O(n^2)$, we have $\mathbb{E}[I_{211}'^{2}]=O(n^{-2}h^{-6})$, and $I_{211}'=O_{p}(n^{-1}h^{-3})$ by the Chebyshev inequality. Hence,
		$I_{211}=O_{p}(n^{-3/2}h^{-3}).$ 
		
		Similarly, we have $I_{212}=O_{p}(n^{-3/2}h^{-4}).$
		
		For $I_{213}$, by triangle inequality,
		\begin{align*}
			&\Vert I_{213}\Vert\leq\frac{\sup\limits_{t\in\mathbb{R}}|K^{(4)}(t)|}{6n^2h^5\inf\limits_{t\in\mathcal{T}}f_o(t)}\sum_{k,j=1}^n\Vert x_j\Vert^3\Vert x_k\Vert\Vert\widehat{\beta}^{(0)}-\beta_0\Vert^{3}\\
			&=\frac{\sup\limits_{t\in\mathbb{R}}|K^{(4)}(t)|}{6h^5\inf\limits_{t\in\mathcal{T}}f_o(t)}\big(\frac{1}{n}\sum_{k=1}^{n}\Vert x_k\Vert^3\big)\big(\frac{1}{n}\sum_{k=1}^{n}\Vert x_k\Vert\big)  \Vert\widehat{\beta}^{(0)}-\beta_0\Vert^{3}\\
			&=O_p\left(\frac{1}{n^{3/2}h^5}\right).
		\end{align*}
		
		For $I_{22}$, we only need to bound its entries:
		$$
		\frac{1}{n}\sum_{k=1}^{n}\frac{\widetilde{f}_h^{(1)}(\varepsilon_k)-f_o^{(1)}(\varepsilon_k)}{f_o(\varepsilon_k)}u_{k-i},~i=1,\cdots,p,
		$$ which is the summation of two terms
		\begin{align*}
			&\frac{1}{n^2h^2}\sum_{j,k=1}^{n}\frac{\!\left\{K^{(1)}\big(\frac{\varepsilon_k-\varepsilon_j}{h}\big)\!-\!\mathbb{E}\left[K^{(1)}\big(\frac{\varepsilon_k-\varepsilon_j}{h}\big)\Big|\varepsilon_k\right]\right\}}{f_o(\varepsilon_k)}u_{k-i}\\
			&\hspace{4mm}+\frac{1}{n^2}\sum_{j,k=1}^{n}\!\frac{\Big[\frac{1}{h^2}\mathbb{E}\left[K^{(1)}\big(\frac{\varepsilon_k-\varepsilon_j}{h}\big)\Big|\varepsilon_k\right]\!-\!f_o^{(1)}(\varepsilon_k)\Big]}{f_o(\varepsilon_k)}u_{k-i}\\
			&\hspace{0mm}\eq I_{221}+I_{222},
		\end{align*}
		The second-order moment of term $I_{221}$ is
		\begingroup\allowdisplaybreaks
		\begin{align*}
			&\mathbb{E}[I_{221}^{2}]=\frac{1}{n^4h^4}\sum_{j_1,j_2,k_1,k_2=1}^{n}\mathbb{E}[u_{k_1-i}u_{k_2-i}]\\
			&\hspace{4mm}\times\mathbb{E}\Bigg[\frac{\!\left\{K^{(1)}\big(\frac{\varepsilon_{k_1}-\varepsilon_{j_1}}{h}\big)\!-\!\mathbb{E}\left[K^{(1)}\big(\frac{\varepsilon_{k_1}-\varepsilon_{j_1}}{h}\big)\Big|\varepsilon_{k_1}\right]\right\}}{f_o(\varepsilon_{k_1})} \frac{\!\left\{K^{(1)}\big(\frac{\varepsilon_{k_2}-\varepsilon_{j_2}}{h}\big)\!-\!\mathbb{E}\left[K^{(1)}\big(\frac{\varepsilon_{k_2}-\varepsilon_{j_2}}{h}\big)\Big|\varepsilon_{k_2}\right]\right\}}{f_o(\varepsilon_{k_2})}\Bigg].
		\end{align*}
		\endgroup
		If $k_1\neq k_2$, then $\mathbb{E}[u_{k_1-i}u_{k_2-i}]=0$ and the corresponding terms are zero. 
		If $k_1=k_2=k$ and $j_1\notin\{j_2,k\}$, 
		\begin{align*}
			\mathbb{E}\Bigg[\frac{\left\{K^{(1)}\big(\frac{\varepsilon_{k}-\varepsilon_{j_1}}{h}\big)-\mathbb{E}\left[K^{(1)}\big(\frac{\varepsilon_{k}-\varepsilon_{j_1}}{h}\big)\Big|\varepsilon_{k}\right]\right\}}{f_o(\varepsilon_{k})}
			\Bigg|\varepsilon_{k}\Bigg]=0,
		\end{align*}
		which implies that the resulting terms are also zero.
		Therefore, there are at most $O(n^2)$ non-zero bounded terms in the summation.
		Accordingly, {$\mathbb{E}[I_{221}^2]=O(n^{-2}h^{-4})$}, then ${I}_{221}=O_{p}(n^{-1}h^{-2})$.
		
		The second-order moment of the term $I_{222}$ is
		\begin{align*}
			&\frac{1}{n^{4}}\!\!\!\sum_{j_1,j_2,k_1,k_2=1}^{n}\mathbb{E}\Bigg[\Bigg(\frac{\frac{1}{h^2}\mathbb{E}\Big[K^{(1)}\!\Big(\frac{\varepsilon_{k_1}-\varepsilon_{j_1}}{h}\Big)\Big|\varepsilon_{k_1}\Big]\!-\!f_o^{(1)}(\varepsilon_{k_1})}{f_o(\varepsilon_{k_1})}\Bigg)\\
			&\times\Bigg(\frac{\frac{1}{h^2}\mathbb{E}\Big[K^{(1)}\Big(\frac{\varepsilon_{k_2}-\varepsilon_{j_2}}{h}\Big)\Big|\varepsilon_{k_2}\Big]\!-\!f_o^{(1)}(\varepsilon_{k_2})}{f_o(\varepsilon_{k_2})}\Bigg)\Bigg]\mathbb{E}[u_{k_1-i}u_{k_2-i}]\\
			&=\frac{1}{n^4}\sum_{j_1,j_2,k=1}^{n}\mathbb{E}\Bigg[\Bigg(\frac{\frac{1}{h^2}\mathbb{E}\Big[K^{(1)}\Big(\frac{\varepsilon_{k}-\varepsilon_{j_1}}{h}\Big)\Big|\varepsilon_{k}\Big]-f_o^{(1)}(\varepsilon_{k})}{f_o(\varepsilon_{k})}\Bigg)\\
			&\times\Bigg(\frac{\frac{1}{h^2}\mathbb{E}\Big[K^{(1)}\Big(\frac{\varepsilon_{k}-\varepsilon_{j_2}}{h}\Big)\Big|\varepsilon_{k}\Big]-f_o^{(1)}(\varepsilon_{k})}{f_o(\varepsilon_{k})}\Bigg)\Bigg]\sigma_u^2.
		\end{align*}
		By Lemma \ref{lem:c4}, we have 
		$$
		\sup_{t\in\mathcal{T}}\Big|\mathbb{E}\Big[\frac{1}{h^2}K^{(1)}\Big(\frac{t-\varepsilon_j}{h}\Big)\Big]-f_o^{(1)}(t)\Big|= O(h^{2}),
		$$
		which derives $\mathbb{E}[I_{222}^2]=O(n^{-1}h^4)$. 
		Thus,  we have $I_{222}=O_p(n^{-1/2}h^{2})$.
		
		For $I_3$, we  write it as
		\begin{align*}
			I_3&=\frac{1}{n}\sum_{k=1}^{n}\frac{(\widehat{f}_h^{(1)}(\varepsilon_k)-f_o^{(1)}(\varepsilon_k))(\widehat{f}_h(\varepsilon_k)-f_o(\varepsilon_k))}{f_o^2(\varepsilon_k)}x_k\\
			&+\frac{1}{n}\sum_{k=1}^{n}\frac{f_o^{(1)}(\varepsilon_k)(\widehat{f}_h(\varepsilon_k)-f_o(\varepsilon_k))}{f_o^2(\varepsilon_k)}x_k\eq I_{31}+I_{32}.
		\end{align*}
		By Lemma \ref{lem:b1}, the term $I_{31}$ is bounded by
		\begin{align*}
			O_p\Big(\sup_{t\in\mathcal{T}}|&\widehat{f}_h^{(1)}(t)-f_o^{(1)}(t)|\sup_{t\in\mathcal{T}}|\widehat{f}_h(t)-f_o(t)|\Big) =O_{p}\Big(h^4+ h^2\sqrt{\frac{\log n}{nh}}+ h^2\sqrt{\frac{\log n}{nh^3}}
			+  \frac{\log n}{nh^2}
			\Big).
		\end{align*} 
		The term $I_{32}$ can be bounded similarly as the term $I_2$, which is $O_{p}(n^{-3/2}h^{-4}+n^{-1}h^{-1}+n^{-1/2}h)$.
		
		Finally, by Lemma  \ref{lem:b1}, we have $I_{4}=O_{p}\Big(h^4+ \frac{\log n}{nh}+h^2\sqrt{\frac{\log n}{nh}}\Big)$. 
		
		By Assumption \ref{asp5}-(iii), all the remainder terms $I_2,I_3,I_4$ are $o_{p}(n^{-1/2})$, which completes our proof.
	\end{proof}
	
	\begin{lem}\label{lem:d2}
		Under Assumptions in Theorem \ref{thm1}, we have
		$
		\widehat{H}_{eM}(\beta^{*})\xrightarrow{p}\Sigma,
		$
		where $\beta^{*}$ lies between $\widehat{\beta}^{eM}$ and $\beta_0$ defined in \eqref{barbeta_em}.
	\end{lem}
	\begin{proof}
		Let  $\mathbb{B}$ be the Euclidean ball centered at $\beta_0$ with radius $\|\beta_0-\widehat{\beta}^{eM}\|\xra{}0$. Thus, $\beta^{*}$ lying between $\widehat{\beta}^{eM}$ and $\beta_0$ must belong to $\mathbb{B}$.
		
		Denote the Hessian matrix of the MLE \eqref{mle} by $H_{eM}(\beta)$, which equals $\widehat{H}_{eM}(\beta)$ except that $\widehat{f}_{h}(\cdot)$ and $\widehat{f}_{h}^{(1)}(\cdot)$ in $\widehat{H}_{eM}$ are replaced by $f_o(\cdot)$ and $f_o^{(1)}(\cdot)$, respectively.
		By the weak law of large numbers for $p$-dependent data \cite{andrews1988}, we have
		$$
		{H}_{eM}(\beta_0)\xrightarrow{p}\Sigma.
		$$
		Therefore, we prove the conclusion by verifying the following uniform convergence
		$$\sup_{\beta\in  \mathbb{B}}
		\|\widehat{H}_{eM}(\beta)-H_{eM}(\beta_0)\|=o_p(1).$$
		Due to the inequality
		\begin{align}\label{hessianconsist}
			\nonumber
			&\sup_{\beta \in  \mathbb{B}}
			\!\|\widehat{H}_{eM}(\beta )-H_{eM}(\beta_0)\|
			\leq\!\sup_{\beta  \in \mathbb{B}}
			\!\|\widehat{H}_{eM}(\beta )-H_{eM}(\beta )\|\\
			&\hspace{23mm}+\sup_{\beta  \in \mathbb{B}}\|H_{eM}(\beta )-H_{eM}(\beta_0)\|,
		\end{align} 
		it suffices to show the uniform convergence of the two terms on the right-hand side of \eqref{hessianconsist}.

		On one hand, by the triangle inequality and Lemma \ref{lem:b1}, for sufficiently large $n$, there holds that
		\begin{align*}
			&\sup_{t\in\mathcal{T}} \Big|\frac{\widehat{f}_h^{(1)}(t)}{\widehat{f}_h(t)}\!-\!\frac{f_o^{(1)}(t)}{f_o(t)}\Big|\leq \sup_{t\in\mathcal{T}} \Big|\frac{\widehat{f}_h^{(1)}(t)\!-\!f_o^{(1)}(t)}{f_o(t)\!-\!|\widehat{f}_h(t)\!-\!f_o(t)|}\Big|\\
			&\hspace{10mm}+\sup_{t\in\mathcal{T}}\Big|\frac{f_o^{(1)}(t)(\widehat{f}_h(t)-f_o(t))}{f_o(t)(f_o(t)-|\widehat{f}_h(t)-f_o(t)|)}\Big|=o_p(1),
		\end{align*}
		under Assumption \ref{asp2}. Moreover, by the identity $a^2-b^2=(a+b)(a-b)$, we obtain
		$$
		\sup_{t\in\mathcal{T}}\Bigg|\Bigg(\frac{\widehat{f}_h^{(1)}(t)}{\widehat{f}_h(t)}\Bigg)^2\!\!-\Bigg(\frac{f_o^{(1)}(t)}{f_o(t)}\Bigg)^2\Bigg|=o_p(1).
		$$
		Similarly, by Theorem 4.1.1 in \cite{rao1983}, we have
		$$
		\sup_{t\in\mathcal{T}}\Bigg|\frac{\widehat{f}_h^{(2)}(t)}{\widehat{f}_h(t)}-\frac{f_o^{(2)}(t)}{f_o(t)}\Bigg|=o_p(1).
		$$
		Therefore, we prove that 
		\begin{align}\label{hem}
			\sup\limits_{\beta\in \mathbb{B} }\|\widehat{H}_{eM}(\beta)-H_{eM}(\beta)\|=o_{p}(1).
		\end{align}

		On the other hand, by the continuity of $H_{eM}(\beta)$ over $\beta$, we achieve
		$$\sup_{\beta  \in \mathbb{B}}\|\widehat{H}_{eM}(\beta )-H_{eM}(\beta_0)\|=o_p(1).$$

		This completes the proof.
	\end{proof}

	\begin{lem}\label{lem:d3}
		Under Assumptions in Theorem \ref{thm2}, we have
		$$
		\sqrt{n}\widehat{S}_{epM}(\beta_0)\xrightarrow[]{d}\mathcal{N}(0,\Sigma).
		$$
	\end{lem}
	\begin{proof}
		By the definition of $\widetilde{f}_{h}(t)$ and $\widetilde{f}^{(1)}_h(t)$ given in \eqref{kde} and \eqref{kde'true}, respectively, we express $\widehat{S}_{epM}(\beta_0)$ given in \eqref{sepm} as  
		\begingroup\allowdisplaybreaks
		\begin{align*}
			\widehat{S}_{epM}(\beta_0)&=\frac{1}{nh}\sum_{k,j=1}^{n}\frac{K^{(1)}\Big(\frac{\varepsilon_k-\varepsilon_j}{h}\Big)}{\sum_{l=1}^{n}K\Big(\frac{\varepsilon_k-\varepsilon_l}{h}\Big)}(x_k-x_j)\\
			&=-\frac{1}{n^2h^2}\sum_{k,j=1}^{n}\frac{K^{(1)}\Big(\frac{\varepsilon_k-\varepsilon_j}{h}\Big)}{\widetilde{f}_{h}(\varepsilon_k)}x_j +\frac{1}{n}\sum_{k=1}^{n}\frac{\widetilde{f}^{(1)}_{h}(\varepsilon_k)}{\widetilde{f}_{h}(\varepsilon_k)}x_k\eq-A_{n}+B_{n}.
		\end{align*}
		\endgroup
		We note that $B_{n}$ can be regarded as $\widehat{S}_{eM}(\beta_0)$ with the initial estimator $\widehat{\beta}^{(0)}$ being the true value $\beta_0$. Similar to the proof of Lemma \ref{lem:d1}, we have
		$$
		\sqrt{n}B_{n}\xrightarrow[]{d}\mathcal{N}(0,\Sigma).
		$$
		Therefore, we only need to prove that $A_{n}=o_{p}(n^{-1/2}).$ 
		
		Since $p$ is finite, it suffices to consider its entries:
		$$
		A_n^i\eq\frac{1}{n^2h^2}\sum_{k,j=1}^{n}\frac{K^{(1)}\Big(\frac{\varepsilon_k-\varepsilon_j}{h}\Big)}{\widetilde{f}_h(\varepsilon_k)}u_{j-i}, ~ i=1,\cdots,p.
		$$
		Applying \eqref{hat-true-2} with $g=\widetilde{f}_h$, we obtain
		\begin{align*}
			\frac{1}{\widetilde{f}_h(t)}&=\frac{1}{f_o(t)}-\frac{\widetilde{f}_h(t)-f_o(t)}{f_o^2(t)}+\frac{(\widetilde{f}_h(t)-f_o(t))^2}{f_o^2(t)\widetilde{f}_h(t)}.
		\end{align*}
		Substituting this into $A_{n}^i$ gives
		\begin{align*}
			{A}_{n}^i&=\frac{1}{n^2h^2} \sum_{k,j=1}^{n} K^{(1)}\Big(\frac{\varepsilon_k-\varepsilon_j}{h}\Big) \frac{u_{j-i}}{f_o(\varepsilon_k)}\\
			&\hspace{4mm}-\frac{1}{n^2h^2}\sum_{k,j=1}^{n} K^{(1)}\Big(\frac{\varepsilon_k-\varepsilon_j}{h}\Big) \frac{\widetilde{f}_h(\varepsilon_k)-f_o(\varepsilon_k)}{f^2_o(\varepsilon_k)}u_{j-i}
			\\
			&\hspace{4mm}+\frac{1}{n^2h^2} \sum_{k,j=1}^{n} K^{(1)}\Big(\frac{\varepsilon_k-\varepsilon_j}{h}\Big)\frac{(\widetilde{f}_h(\varepsilon_k)-f_o(\varepsilon_k))^2}{f_o^2(\varepsilon_k)\widetilde{f}_{h}(\varepsilon_k )}u_{j-i}\\
			&\eq {A}_{n1}^i-{A}_{n2}^i+{A}_{n3}^i.
		\end{align*}
		
		For ${A}_{n1}^i$,  its second moment is
		\allowdisplaybreaks
		\begin{align*}
			\mathbb{E}\big((A_{n1}^i)^2\big)&=\frac{1}{n^4h^4}\sum_{k_1,k_2,j_1,j_2=1}^{n}\mathbb{E}[u_{j_1-i}u_{j_2-i}]  \mathbb{E}\Bigg[\frac{K^{(1)}\left(\frac{\varepsilon_{k_1}-\varepsilon_{j_1}}{h}\right)K^{(1)}\left(\frac{\varepsilon_{k_2}-\varepsilon_{j_2}}{h}\right)}{f_o(\varepsilon_{k_1})f_o(\varepsilon_{k_2})}\Bigg]\\
			&\hspace{0.8mm}=\frac{\sigma_u^2}{n^4h^4}\sum_{k_1,k_2,j=1}^{n}\mathbb{E}\Bigg[\frac{K^{(1)}\left(\frac{\varepsilon_{k_1}-\varepsilon_{j}}{h}\right)K^{(1)}\left(\frac{\varepsilon_{k_2}-\varepsilon_{j}}{h}\right)}{f_o(\varepsilon_{k_1})f_o(\varepsilon_{k_2})}\Bigg].
		\end{align*}
		Note that
		\allowdisplaybreaks
		\begin{align*}
			 \mathbb{E}\Bigg[\frac{K^{(1)}\left(\frac{\varepsilon_{k_1}-\varepsilon_{j}}{h}\right)K^{(1)}\left(\frac{\varepsilon_{k_2}-\varepsilon_{j}}{h}\right)}{f_o(\varepsilon_{k_1})f_o(\varepsilon_{k_2})}\Bigg] =\!\int\!K^{(1)}\Big(\frac{s_{k_1}\!-\!s_{j}}{h}\Big)K^{(1)}\Big(\frac{s_{k_2}\!-\!s_{j}}{h}\Big)f_o(s_{j})\d s_{j}\d s_{k_1}\d  s_{k_2}. 
		\end{align*}
		If $k_1\neq k_2$, then the expectation above is zero by the symmetry of $K(\cdot)$  for all $j=1,\cdots,n$.
		Consequently, we have
		$$
		\mathbb{E}\big((A_{n1}^i\big)^2)=\frac{\sigma_u^2}{n^4h^4}\sum_{k,j=1}^{n}\mathbb{E}\bigg(
		\frac{K^{(1)} 
			\big(\frac{\varepsilon_{k}-\varepsilon_{j}}{h}\big)  }{f_o(\varepsilon_{k})}\bigg)^2.
		$$
		By Assumption \ref{asp2}-(iii)  and Assumption \ref{asp5}-(ii), we have 
		$$\mathbb{E}\Big[K^{(1)} \Big(\frac{\varepsilon_{k}-\varepsilon_{j}}{h}\Big)^2\big/f_o^2(\varepsilon_{k})\Big]<\infty,$$ 
		which implies that  $\mathbb{E}\big((A_{n1}^i\big)^2)=O(n^{-2}h^{-4})$ and further $A_{n1}^i=O_{p}(n^{-1}h^{-2})$.
		
		For $A_{n2}^i$, we plug in the expansion of $\widetilde{f}_h(\cdot)$ to obtain
		\begingroup\allowdisplaybreaks
		\begin{align*}
			A_{n2}^i&=\frac{1}{n^3h^2}\sum_{k,j,l=1}^{n}K^{(1)}\Big(\frac{\varepsilon_k-\varepsilon_j}{h}\Big)\frac{u_{j-i}}{f_o^2(\varepsilon_k)} \left(\frac{1}{h}K\Big(\frac{\varepsilon_k-\varepsilon_l}{h}\Big)-f_o(\varepsilon_k)\right)\\
			&=\frac{1}{n^3h^2}\sum_{k,j,l=1}^{n}K^{(1)}\Big(\frac{\varepsilon_k-\varepsilon_j}{h}\Big)\frac{u_{j-i}}{f_o^2(\varepsilon_k)} \left(\frac{1}{h}K\Big(\frac{\varepsilon_k-\varepsilon_l}{h}\Big)-\mathbb{E}\Big[\frac{1}{h}K\Big(\frac{\varepsilon_k-\varepsilon_l}{h}\Big)\Big|\varepsilon_k\Big]\right)\\
			&\hspace{4mm}+\frac{1}{n^3h^2}\sum_{k,j,l=1}^{n}K^{(1)}\Big(\frac{\varepsilon_k-\varepsilon_j}{h}\Big)\frac{u_{j-i}}{f_o^2(\varepsilon_k)} \left(\mathbb{E}\Big[\frac{1}{h}K\Big(\frac{\varepsilon_k-\varepsilon_l}{h}\Big)\Big|\varepsilon_k\Big]-f_o(\varepsilon_k)\right)\\
			&\eq A_{n21}^i+A_{n22}^i.
		\end{align*}
		\endgroup
		For simplicity, we adopt the notation
		$$
		K_e(h,x,y)=\frac{1}{h}K\Big(\frac{x-y}{h}\Big)-\frac{1}{h}\mathbb{E}\Big[K\Big(\frac{x-y}{h}\Big)\Big|y\Big].
		$$
		The second absolute moment of $A_{n21}^i$ is
		\begin{align*}
			&\frac{1}{n^6h^4}\sum_{\substack{k_1,k_2,j_1,\\j_2,l_1,l_2=1}}^{n}\mathbb{E}\Bigg[\frac{K^{(1)}\Big(\frac{\varepsilon_{k_1}-\varepsilon_{j_1}}{h}\Big)K^{(1)}\Big(\frac{\varepsilon_{k_2}-\varepsilon_{j_2}}{h}\Big)}{f_o^2(\varepsilon_{k_1})f_o^{2}(\varepsilon_{k_2})} K_e(h,\varepsilon_{k_1},\varepsilon_{l_1})K_e(h,\varepsilon_{k_2},\varepsilon_{l_2})\Bigg] \mathbb{E}[u_{j_1-i}u_{j_2-i}]
			\\&=\frac{\sigma_u^2}{n^6h^4}\sum_{\substack{k_1,k_2,j,\\l_1,l_2=1}}^{n}\mathbb{E}\Bigg[\frac{K^{(1)}\Big(\frac{\varepsilon_{k_1}-\varepsilon_{j}}{h}\Big)K^{(1)}\Big(\frac{\varepsilon_{k_2}-\varepsilon_{j}}{h}\Big)}{f_o^2(\varepsilon_{k_1})f_o^{2}(\varepsilon_{k_2})} K_e(h,\varepsilon_{k_1},\varepsilon_{l_1})K_e(h,\varepsilon_{k_2},\varepsilon_{l_2})\Bigg].
		\end{align*}
		For $l_1\notin\{k_1,k_2,j,l_2\}$, there holds that
		$$\mathbb{E}\Big[K_e(h,\varepsilon_{k_1},\varepsilon_{l_1})\Big|\varepsilon_{k_1},\varepsilon_{k_2},\varepsilon_{j},\varepsilon_{l_2}\Big]=0,$$
		and these terms in the summation vanish.
		Therefore, at most $O(n^4)$ terms are non-zero and each one is bounded by
		$$
		\frac{4\Big[\sup\limits_{t\in\mathbb{R}}|K^{(1)}(t)|\Big]^2\Big[\sup\limits_{t\in\mathbb{R}}|K(t)|\Big]^2}{h^2\big[\inf\limits_{t\in\mathcal{T}}f_o(t)\big]^4}=O\Big(
		\frac{1}{h^2}
		\Big),
		$$
		which implies that $\mathbb{E}[A_{n21}^i]^2=O(n^{-2}h^{-6})$ and further $A_{n21}^i=O_{p}(n^{-1}h^{-3})$.
		
		For $A_{n22}^i$, by Assumption \ref{asp1} and Lemma \ref{lem:c4} 
		\begingroup\allowdisplaybreaks
		\begin{align*}
			|A_{n22}^i|&\leq\frac{1}{n^3h^2}\sum_{k,j,l=1}^{n}\Big|K^{(1)}\Big(\frac{\varepsilon_k-\varepsilon_j}{h}\Big)\Big|\frac{|u_{j-i}|}{f_o^2(\varepsilon_k)} \Bigg|\mathbb{E}\Big[\frac{1}{h}K\Big(\frac{\varepsilon_k-\varepsilon_l}{h}\Big)\Big|\varepsilon_k\Big]-f_o(\varepsilon_k)\Bigg|\\
			& \leq\frac{\sup\limits_{t\in\mathbb{R}}|K^{(1)}(t)|}{\inf\limits_{t\in\mathcal{T}}f_o^2(t)}\sup_{t\in\mathcal{T}}\Bigg|\mathbb{E}\Big[\frac{1}{h}K\Big(\frac{t-\varepsilon_1}{h}\Big)\Big]-f_o(t)\Bigg| \frac{1}{n^3h^2}\sum_{k,j,l=1}^{n}|u_{j-i}|\\
			&=O\big(
			h^{2q} 
			\big)
			O_p\Big(\frac{1}{ h^2}
			\Big)
			O_p\Big(\frac{1}{n}\sum_{j=1}^{n}|u_{j-i}|
			\Big) =O_p\Big(
			h^{2q-2} 
			\Big).
		\end{align*}
		\endgroup
		
		For $A_{n3}^i$, there holds that
		\begin{align*}
			|A_{n3}^i|&\leq\frac{1}{n^2h^2}\!\sup_{t\in\mathcal{T}}\!\frac{|\widetilde{f}_h(t)\!-\!f_o(t)|^2}{f_o^2(t)\widetilde{f}_{h}(t)} 
			\!\!\sum_{k,j=1}^{n} \!\Big|K^{(1)}\Big(\frac{\varepsilon_k\!-\!\varepsilon_j}{h}\Big)\Big| |u_{j-i}|\\
			&\leq\frac{1}{h^2}\sup_{t\in\mathcal{T}}\frac{|\widetilde{f}_h(t)\!-\!f_o(t)|^2}{f_o^2(t)\widetilde{f}_{h}(t)}  \sup\limits_{t\in\mathbb{R}}|K^{(1)}(t)| \Big(\frac{1}{n} \sum_{j=1}^{n} |u_{j-i}|\Big)\\
			&=O\Big(\frac{1}{h^2}\Big)O_{p}\Big(h^{4q}+\frac{\log n}{nh}+h^{2q}\sqrt{\frac{\log n}{nh}}\Big)O_{p}(1)
			\\
			&=O_{p}\Big(h^{4q-2}+\frac{\log n}{nh^3}+h^{2q-2}\sqrt{\frac{\log n}{nh}}\Big),
		\end{align*}
		where   
		\allowdisplaybreaks
		\begin{align*}
			\sup_{t\in\mathcal{T}}\frac{|\widetilde{f}_h(t)-f_o(t)|^2}{f_o^2(t)\widetilde{f}_{h}(t)} & \leq\frac{\Big(\sup\limits_{t\in\mathcal{T}}|\widetilde{f}_h(t)-f_o(t)|\Big)^2}{\Big(\inf\limits_{t\in\mathcal{T}}f_o(t)\Big)^2\Big[\inf\limits_{t\in\mathcal{T}}f_o(t)-\sup\limits_{t\in\mathcal{T}}|\widetilde{f}_h(t)-f_o(t)|\Big]}\\
			&\leq\frac{\Big(\sup\limits_{t\in\mathcal{T}}|\widetilde{f}_h(t)-f_o(t)|\Big)^2}{\Big(\inf\limits_{t\in\mathcal{T}}f_o(t)\Big)^3}=O_{p}\Big(h^{4q}+\frac{\log n}{nh}+h^{2q}\sqrt{\frac{\log n}{nh}}\Big)
		\end{align*}
		via \eqref{kdeuniform} for a $q$-th order kernel.
		
		Combining the rates of the terms ${A}_{n1}^i$, ${A}_{n2}^i$, and ${A}_{n3}^i$ derived above, by Assumption \ref{asp5}-(iii), we reach $A_{n}=o_{p}(n^{-1/2}).$
	\end{proof}

	\renewcommand{\theequation}{C\arabic{equation}}
	\setcounter{equation}{0}
	\setcounter{thm}{0}
	\renewcommand{\thethm}{C\arabic{thm}}
	\setcounter{lem}{0}
	\renewcommand{\thelem}{C\arabic{lem}}
	\setcounter{rmk}{0}
	\renewcommand{\thermk}{C\arabic{rem}}
	\setcounter{subsection}{0}
	\renewcommand{\thesection}{C}
	
	\renewcommand{\thesection}{C}
	
	\section*{Appendix C: Uniform convergence of estimated density functions}
	
	This subsection establishes the uniform convergence rates of the kernel density estimator based on the estimated noise samples.
	
	\begin{lem}\label{lem:b1}
		Suppose that Assumptions \ref{asp1}--\ref{asp2} and \ref{asp4}--\ref{asp5} hold. Then, it follows that
		\allowdisplaybreaks
		\begin{align}
			\label{qq}
			&\sup_{t\in\mathcal{T}}  |\widehat{f}_{h}(t)-f_o(t)| = O_p\Big(h^2 + \sqrt{\frac{\log n}{nh}} \Big),\\
			\label{qq2}
			&\sup_{t\in\mathcal{T}}  |\widehat{f}^{(1)}_{h} (t) -f_o^{(1)} (t)| =O_p\Big(h^2+\sqrt{\frac{\log n}{nh^3}} \Big).
		\end{align}
		If Assumption \ref{asp6} additionally holds, then the terms \( h^2 \) on the right-hand side above are improved to \( h^{2q} \).
	\end{lem}
	\begin{proof}
		We start with the decomposition: 
		$$
		\widehat{f}_{h}^{(r)}(t)-f_o^{(r)}(t)=\widehat{f}_{h}^{(r)}(t)-\widetilde{f}_{h}^{(r)}(t)+\widetilde{f}_{h}^{(r)}(t)-f_o^{(r)}(t)
		$$
		for $r=0,1$, 
		where $\widetilde{f}_{h}(t)$ and $\widetilde{f}_{h}^{(1)}(t)$  
		are defined by \eqref{kde} and \eqref{kde'true}, respectively.
		
		For $\widetilde{f}_{h}^{(r)}(t)-f_o^{(r)}(t)$, by the uniform convergence rate from \cite{silverman1978,li2007} and Lemma \ref{lem:c4}, we have
		\begin{align}\label{kdeuniform}
			\sup_{t\in\mathcal{T}}|\widetilde{f}_{h}^{(r)}(t)-f_o^{(r)}(t)|=O_p\Big(h^2+\sqrt{\frac{\log n}{nh^{2r+1}}}\Big).
		\end{align}
		Next, we consider $\widehat{f}_{h}^{(r)}(t)-\widetilde{f}_h^{(r)}(t)$, and we take $r=0$ for illustration.
		It follows from  Taylor expansion that
		\begin{align*}
			\widehat{f}_{h}(t)-\widetilde{f}_{h}(t)&=\frac{1}{nh^2}\sum_{k=1}^{n}K^{(1)}\Big(\frac{t-\varepsilon_k}{h}\Big)x_k^T(\widehat{\beta}^{(0)}-\beta_0)\\
			&\hspace{5mm}+\frac{1}{2nh^3}\sum_{k=1}^{n}K^{(2)}\Big(\frac{t-\varepsilon_k}{h}\Big)(x_k^T(\widehat{\beta}^{(0)}-\beta_0))^2\\
			&\hspace{5mm}+\frac{1}{6nh^4}\sum_{k=1}^{n}K^{(3)}\Big(\frac{t-\bar{\varepsilon}_k}{h}\Big)(x_k^T(\widehat{\beta}^{(0)}-\beta_0))^3\\
			&\hspace{-1mm}\eq J_1+J_2+J_3,
		\end{align*}
		where $\bar{\varepsilon}_k$ lies between $\varepsilon_k$ and $\widehat{\varepsilon}_{k}$.
		
		For $J_1$, we write it as the summation
		$$
		J_1=\frac{1}{\sqrt{n}h^2}\sum_{j=1}^{p}(\widehat{\beta}^{(0)}-\beta_0)_j
		\bigg(\!
		\underbrace{\frac{1}{\sqrt{n}}\sum_{k=1}^{n}K^{(1)}\Big(\frac{t-\varepsilon_k}{h}\Big)u_{k-j}}_{J_{1j}(t)}\!\bigg).
		$$
		We next   employ the empirical process to derive a bound for $\sup\limits_{t\in\mathcal{T}}|J_{1j}(t)|$.
		By Lemma \ref{lem:c5}, we have 
		$$
		\mathbb{E}\Big(\sup_{t\in\mathcal{T}} |J_{1j}(t)|\Big)<\infty,
		$$
		which yields 
		$
		\sup_{t\in\mathcal{T}}|J_{1j}(t)|=O_{p}(1)
		$ by the Markov inequality.
		Consequently, $$\sup_{t\in\mathcal{T}}|J_{1}|=O_{p}
		\Big(
		\frac{1}{\sqrt{n}h^2}
		\Big)O_{p}
		\Big(
		\frac{1}{\sqrt{n} }
		\Big)O_p(1)=O_{p}
		\Big(
		\frac{1}{ nh^2}
		\Big).
		$$ 
		
		For $J_2$,  consider the matrix
		$$
		C_{n}\eq\frac{1}{n}\sum_{k=1}^{n}K^{(2)}\Big(\frac{t-\varepsilon_k}{h}\Big)x_kx_k^T,
		$$
		whose $(i,j)$-th element $(C_n)_{ij}$ 
		satisfies
		\begin{align*}
			\mathbb{E}\Big[\sup_{t\in\mathcal{T}}|(C_n)_{ij}|\Big]
			&=
			\mathbb{E}\Big[\sup_{t\in\mathcal{T}}\Big|\frac{1}{n}\sum_{k=1}^{n}K^{(2)}\Big(\frac{t-\varepsilon_k}{h}\Big)u_{k-i}u_{k-j}\Big|\Big]\\
			&\leq\Big(\sup_{x\in\mathbb{R}}
			|K^{(2)} (x)|\Big)
			\mathbb{E}|u_{k-i}u_{k-j}|
			<\infty.
		\end{align*}
		This means that $\sup_{t\in\mathcal{T}}|(C_{n})_{i,j}|=O_{p}(1)$ for $i,j=1,\cdots,p$
		by the Markov inequality.
		Therefore, we acquire $$\sup_{t\in\mathcal{T}}|J_2| 
		\leq 
		\frac{1}{2h^{3}}
		\|\beta_0-\widehat{\beta}^{(0)}\|^2
		\sup_{t\in\mathcal{T}}\|C_{n}\|= O_p\Big(\frac{1}{nh^{3}}\Big).$$
		
		For $J_3$, by Assumption \ref{asp5}, there holds that
		\allowdisplaybreaks
		\begin{align*}
			\sup_{t\in\mathcal{T}}|J_3|&\leq\frac{\sup\limits_{x\in\mathbb{R}}|K^{(3)}(x)|}{6nh^4}\sum_{k=1}^{n}\Vert x_k\Vert^{3}\Vert\widehat{\beta}^{(0)}-\beta_0\Vert^3\\
			&=\frac{\sup\limits_{x\in\mathbb{R}}|K^{(3)}(x)|}{6h^4}
			\Big(\frac1n\sum_{k=1}^{n}
			\Vert x_k\Vert^{3}
			\Big)
			\Vert\widehat{\beta}^{(0)}-\beta_0\Vert^3\\
			&=O_p
			\Big(
			\frac{1}{h^4} 
			\Big)
			O_p(1)
			O_{p}\Big(
			\frac{1}{n^{3/2}} 
			\Big)=O_p\left(\frac{1}{n^{3/2}h^4}\right).
		\end{align*}
		Therefore, under Assumption \ref{asp5}-(iii)
		\begin{align*}
			\sup_{t\in\mathcal{T}} | &\widehat{f}_{h}(t)-\widetilde{f}_{h}(t)|
			\leq \sup_{t\in\mathcal{T}} | J_1| +\sup_{t\in\mathcal{T}} | J_2| +\sup_{t\in\mathcal{T}} | J_3| \\
			&=O_p\left(\frac{1}{nh^2}\right)
			+O_p\left(\frac{1}{n h^3}\right)
			+O_p\left(\frac{1}{n^{3/2}h^4}\right)\\
			&=o_p\Big(\sqrt{\frac{\log n}{nh}}\Big),
		\end{align*}
		which implies \eqref{qq} via \eqref{kdeuniform}.
		
		Similarly, we can prove  
		\begin{align*}
			\sup_{t\in\mathcal{T}}|\widehat{f}_{h}^{(1)}(t)\!-\!\widetilde{f}_{h}^{(1)}(t)|&
			=O_{p}(n^{-1}\!h^{-3}\!\!+\!n^{-3/2}h^{-4}\!\!+\!n^{-3/2}h^{-5})\\
			&=o_p\Big(\sqrt{\frac{\log n}{nh^{3}}}\Big),
		\end{align*} 
		and obtain \eqref{qq2}.
		
		For the $q$-th order kernel, we instead bound the bias term $h^2$ in \eqref{kdeuniform} by $h^{2q}$ using Lemma \ref{lem:c4}.
	\end{proof}
	
	\begin{lem}\label{lem:b2}
		Suppose that the kernel function $K(\cdot)$ satisfies Assumption \ref{asp5}. If $f_o(\cdot)$ is uniformly continuous, then the KDE \eqref{fbetat} converges to  $f(t,\beta)$ given in \eqref{ftbeta} uniformly over $\beta$, i.e.,
		$$
		\sup_{t\in\mathcal{T},\beta\in\Theta}|\widehat{f}_h(t,\beta)-f(t,\beta)|\xrightarrow{p}0.
		$$
	\end{lem}
	\begin{proof}
		This result is a direct corollary of Theorem A in \cite{silverman1978}.
	\end{proof}
	
	\begin{lem}\label{lem:c4}
		\cite[Theorem 4.1.2]{rao1983}
		Suppose that the density function $f_o^{(2q)}$ exists and is bounded.
		Then for the $q$-th order kernel $K(\cdot)$, there holds that
		\begin{align*}
			&\sup_{t\in\mathcal{T}}\Big|\mathbb{E}\Big[\frac{1}{h}K\Big(\frac{t-\varepsilon_1}{h}\Big)\Big]-f_o(t)\Big|  =O(h^{2q}).
		\end{align*}
		If further $f_o^{(2q+1)}$ exists and is bounded, then
		\begin{align*}
			&\sup_{t\in\mathcal{T}}\Big|\mathbb{E}\Big[\frac{1}{h^2}K^{(1)}\Big(\frac{t-\varepsilon_1}{h}\Big)\Big]-f_o^{(1)}(t)\Big|  =O(h^{2q}).
		\end{align*}
	\end{lem}

	\renewcommand{\theequation}{D\arabic{equation}}
	\setcounter{equation}{0}
	\setcounter{thm}{0}
	\renewcommand{\thethm}{D\arabic{thm}}
	\setcounter{lem}{0}
	\renewcommand{\thelem}{D\arabic{lem}}
	\setcounter{rmk}{0}
	\renewcommand{\thermk}{D\arabic{rem}}
	\setcounter{subsection}{0}
	\renewcommand{\thesection}{D}

	\section*{Appendix D: Auxiliary lemmas}
	In this part, we present auxiliary lemmas used in the proofs of theorems. 
	We first introduce an inequality and an equality in the proofs of theorems. 
	\begin{lem}\label{lem:c0}
		Let $a(t)$ and $b(t)$ be functions with $a(t)>|b(t)|>0$ for all $t\in\mathcal{T}$.
		Thus, we have
		$$
		\sup_{t\in\mathcal{T}}\Big|\log\left(1+\frac{b(t)}{a(t)}\right)\Big|\leq\frac{\sup\limits_{t\in\mathcal{T}}|b(t)|}{\inf\limits_{t\in\mathcal{T}}a(t)-\sup\limits_{t\in\mathcal{T}}|b(t)|}.
		$$
	\end{lem}
	\begin{proof}
		If $b(t)\geq0$, then $0\le\log(1+b/a)\le b/a\le |b|/(a-|b|)$; if $b(t)<0$, then $0\le-\log(1+b/a)=\log\frac{a}{a+b}\le\frac{a}{a+b}-1=\frac{|b|}{a-|b|}$. Taking the supremum over $t\in\mathcal{T}$ completes the proof.
	\end{proof}
	
	\begin{lem}\label{lem:c-1}
		For any function $g(t)$ and the noise density function $f_o(t)$, we have the identity
		\begin{align}
			\label{hat-true-2}
			\frac{1}{g(t)}&=\frac{1}{f_o(t)}-\frac{g(t)-f_o(t)}{f_o^2(t)}+\frac{(g(t)-f_o(t))^2}{f_o^2(t)g(t)}.
		\end{align}
	\end{lem}
	\begin{proof}
		This identity follows by direct computation.
	\end{proof}
	
	We next state  lemmas that establish consistency and asymptotic normality of the proposed nonparametric MLEs.
	
	\begin{lem}
		\cite[Theorem 1]{potscher1989} \label{lem:c2}
		Suppose the sequence $\{z_k\}_{k=1}^{n}$ is $p$-dependent and identically distributed, the function $q(z,\beta)$ is continuous in $(z,\beta)$,
		and
		$\mathbb{E}[\sup_{\beta\in\Theta}|q(z_1,\beta)|^{1+\delta}]<\infty$ for some $\delta>0$, where $\Theta$ is a compact set. 
		If $\frac{1}{n}\sum_{k=1}^{n}q(z_k,\beta)$ converges in probability pointwise to $\mathbb{E}[q(z_1,\beta)]$ for all $\beta\in\Theta$, then
		$$
		\sup_{\beta\in\Theta}\Bigg|\frac{1}{n}\sum_{k=1}^{n} 
		q(z_k,\beta)-\mathbb{E}[q(z_k,\beta)]
		\Bigg|
		\xrightarrow[]{p} 0.
		$$
	\end{lem}
	
	\begin{lem}
		\cite[Theorem 5.7]{van2000}\label{lem:c1}
		Let $M_n(\beta)$ be a sequence of random functions and let $M(\beta)$ be a deterministic function that is well-separated at $\beta_0$. 
		If
		$
		\sup_{\beta\in\Theta}\left|M_n(\beta)-M(\beta)\right|\xrightarrow[]{p} 0
		$, where $\Theta$ is a compact set,
		then the maximizer $\widehat{\beta}_n$ of $M_n(\beta)$ converges in probability to $\beta_0$.
	\end{lem}

	\begin{lem}\cite[Lemma 2.8]{van2000} \label{lem:c3}
		Let $Y_n$ be a sequence of random vectors, and let $X_n$ be a sequence of random matrices.
		If $X_n$ 
		converges in probability to a positive definite matrix $X$ and
		$Y_n$ converges in distribution to a random vector $Y$, then
		$$
		X_n^{-1}Y_n\xrightarrow{d}X^{-1}Y.
		$$
	\end{lem}

	We provide the uniform bound on the term $J_{1}$ in Lemma \ref{lem:b1} based on empirical process that involves Vapnik–Chervonenkis (VC) theory    and maximum inequality \cite{anthony2009,wellner2023}.

	\begin{lem}\label{lem:c5}
		
		Suppose Assumptions \ref{asp1}--\ref{asp2} holds and $K^{(1)}(t)$ is a bounded variation function. Then, it takes place
		\begin{align*}
			&\mathbb{E}\bigg(\sup_{t\in\mathcal{T}}\Big|\frac{1}{\sqrt{n}}\sum_{k=1}^{n}K^{(1)}\Big(\frac{t-\varepsilon_k}{h}\Big)u_{k-j}\Big|\bigg)<\infty.
		\end{align*}
	\end{lem}
	
	\begin{proof}
		Define the function class 
		$$
		\mathcal{F} \eq\Big\{f(x,u)\eq K^{(1)}\Big(\frac{t-x}{h}\Big)u\Big|h>0,t\in\mathcal{T}\Big\}.
		$$ 
		Since $K^{(1)}(t)$ is a bounded variation function,   it can be expressed as the difference of two increasing functions, i.e., $K^{(1)}_1(t)-K^{(1)}_2(t)$  \cite[Theorem 3.3]{stein2005}.
		Therefore, we define two base function classes $\mathcal{F}_{s}$ for $s=1,2$ 
		$$
		\mathcal{F}_{s}\eq\Big\{K^{(1)}_s\Big(\frac{t-x}{h}\Big)\Big| h>0,t\in\mathcal{T}\Big\},
		$$
		which consists of scaling and translation of a monotone function. 
		By Lemma 2.6.17 in \cite{wellner2023}, the VC-dimension of $\mathcal{F}_{s}$ is finite. 
		Moreover,
		by Lemma 2.6.20 (\romannumeral4,\romannumeral5) in \cite{wellner2023}, the function class defined by
		\begin{align*}
			\mathcal{F}'\eq\Big\{
			K^{(1)}_1\Big(\frac{t-x}{h}\Big)
			-K^{(1)}_2\Big(\frac{t-x}{h}\Big)
			\Big|h>0,t\in\mathcal{T}\Big\},
		\end{align*}
		is a VC class with finite dimension.
		Note that all elements of the function class $\mathcal{F}$ are produced by multiplying each function of $\mathcal{F}'$ with the linear function $g(u)=u$.
		Therefore, $\mathcal{F}$ is also a VC class with finite dimension by Lemma 2.6.20 (\romannumeral6) in \cite{wellner2023}.
		
		Note that for any $f\in\mathcal{F} $,
		$$
		|f(x,u)|\leq\Big(\sup_{x\in\mathbb{R}}|K^{(1)}(x)|\Big)\times |u|\eq F(u).
		$$
		The sequence  $\{[\varepsilon_k,u_{k-j}]^T\}$ consists of i.i.d. random vectors drawn from some measure $P$.
		Therefore, by \cite[Theorem 2.14.1]{wellner2023},
		we have
		\begin{align*}
			&\mathbb{E}\bigg(\sup_{t\in\mathcal{T}}\Big|\frac{1}{\sqrt{n}}\sum_{k=1}^{n}K^{(1)}\Big(\frac{t-\varepsilon_k}{h}\Big)u_{k-j}\Big|\bigg)\\
			&\leq
			C\|F\|_{L_{2}(P)}
			\sup_{Q}
			\!\int_{0}^{1}\!\!\sqrt{1+\log N(\epsilon\|F\|_{L_2(Q)},\mathcal{F},L_{2}(Q))}~\d\epsilon,
		\end{align*}
		where $C$ is a constant, $\|F\|_{L^2(\mu)}$ denotes the $L_2$-norm of the function $F(u)$ with respect to probability measure $\mu$, $Q$ takes over all probability measures,  and $N(\epsilon,\mathcal{F},\mu)$ is the $\epsilon$-covering number of a function family $\mathcal{F}$ under measure $\mu$ (see \cite{kosorok2007} for its definition). 
		
		It is clear that 
		$$\|F\|_{L_{2}(P)}
		=\Big(\int F(u)^2 \d P\Big)^\frac{1}{2}
		\leq
		\Big(\sup\limits_{x\in\mathbb{R}}|K^{(1)}(x)|\Big)
		\sigma_{u}<\infty.$$
		
		By  \cite[Theorem 2.6.7]{wellner2023}, there holds that
		$$
		\sup_{Q}N(\epsilon\|F\|_{L_2(Q)},\mathcal{F},L_{2}(Q))\leq C\Big(\frac{1}{\epsilon}\Big)^{2V(\mathcal{F})},\ \epsilon\in(0,1),
		$$
		where $V(\mathcal{F})$ is the VC-dimension of $\mathcal{F}$, which is finite, and $C$ is a  constant that depends on $V(\mathcal{F})$.
		Thus, the integral
		\begin{align*}
			&\sup_{Q}
			\!\int_{0}^{1}\!\!\sqrt{1+\log N(\epsilon\|F\|_{L_2(Q)},\mathcal{F},L_{2}(Q))}~\d\epsilon\\
			&\hspace{25.5mm}\leq \int_{0}^{1}\sqrt{1+C_0\log(1/\epsilon)}~\d\epsilon<\infty,
		\end{align*}
		where $C_0$ is a  constant.
		
		Combining the results above, we complete the proof.
	\end{proof}

\bibliographystyle{unsrt}
	\bibliography{ref}

\end{document}